\documentclass[prb,twocolumn,longbibliography,amsmath,superscriptaddress,amssymb,nobibnotes]{revtex4-2}

\usepackage{graphicx}
\usepackage{dcolumn}
\usepackage{bm}
\usepackage{amssymb}
\usepackage{amsmath}
\usepackage{color}
\usepackage[dvipsnames]{xcolor}
\usepackage{placeins}
\usepackage{epstopdf}
\usepackage{multirow}
\usepackage[normalem]{ulem}
\usepackage[linkcolor=blue,urlcolor=blue,citecolor=blue,colorlinks=true]{hyperref}

\begin{document}

\title{
Single crystal growth and physical characterization to fine tune YbIn$_{1-x}$T$_x$Cu$_4$ (T = Au, Ag) towards the critical endpoint of the valence transition }

\author{Michelle Ocker}
\affiliation{
 Physikalisches Institut, 
Goethe-Universit\"at Frankfurt, 60438 Frankfurt/M, Germany
}
\author{Bereket Ghebretinsae }
\affiliation{
 Physikalisches Institut, 
Goethe-Universit\"at Frankfurt, 60438 Frankfurt/M, Germany
}
\author{Jan-Niklas Zimmermann }
\affiliation{
 Physikalisches Institut, 
Goethe-Universit\"at Frankfurt, 60438 Frankfurt/M, Germany
}
\author{Sophie Würtele}
\affiliation{
 Physikalisches Institut, 
Goethe-Universit\"at Frankfurt, 60438 Frankfurt/M, Germany
}

\author{Bernd Wolf }
\affiliation{
 Physikalisches Institut, 
Goethe-Universit\"at Frankfurt, 60438 Frankfurt/M, Germany
}
\author{Alexandr Virovets}
\affiliation{
 Institute of Inorganic Chemistry, 
Goethe-Universit\"at Frankfurt, 60438 Frankfurt/M, Germany
}
\author{Michael Lang }
\affiliation{
Physikalisches Institut, 
Goethe-Universit\"at Frankfurt, 60438 Frankfurt/M, Germany
}
\author{Kristin Kliemt}
\email{kliemt@physik.uni-frankfurt.de}
\affiliation{
 Physikalisches Institut, 
Goethe-Universit\"at Frankfurt, 60438 Frankfurt/M, Germany
}
\author{Cornelius Krellner}
\affiliation{
 Physikalisches Institut, 
Goethe-Universit\"at Frankfurt, 60438 Frankfurt/M, Germany
}

\date{\today}
\begin{abstract}
\noindent
Pure as well as Ag- and Au-substituted YbInCu$_4$ single crystals were structurally and chemically characterized and investigated by means of heat capacity, magnetization, resistivity and ultrasonic measurements. 
We studied the influence of different compositions of the initial melt as well as of Au and Ag substitutions on the valence change and investigated whether this change occurs via a first-order phase transition or via crossover. We constructed a phase diagram of YbInCu$_4$ as a function of various substitutions and show that the position of the critical endpoint of the valence transition depends on the substituent and on the conditions under which the samples were grown. Multiple thermal cycles through the first-order transition lead to a significant modification of the physical properties which clearly demonstrated the influence of defects in substituted YbInCu$_4$.
\end{abstract}

\maketitle

\section{Introduction}\noindent
Nowadays, the investigation of quantum materials subject to elastic tuning and the resulting electronic orders has become a central part in the research on strongly correlated electron systems, leading to discoveries such as novel states of matter like nematic quantum liquids  \cite{bohmer2016electronic, yao2022electronic}, 
unique nonlinear electron-lattice interactions in unconventional superconductors \cite{li2022elastocaloric}, 
pseudoelasticity \cite{Xiao2021}, and critical elasticity \cite{Gati2016}. 
Likewise, particularly interesting effects, resulting from a strong coupling of correlated electrons to the elastic degrees of freedom, can be expected close to second-order critical endpoints (CEPs) which are amenable to pressure tuning. For example, clear indications for the phenomenon of "critical elasticity" were found at the CEP of the Mott metal-insulator in an organic charge-transfer salt, indicating an intimate coupling of the crystal lattice to the critical electronic system \cite{Gati2016}. In addition to the lattice- and charge degrees of freedom in the Mott scenario, the involvement of magnetic degrees of freedom may pose an interesting extension of this concept. Rare earth compounds, in which the magnetic properties are determined by the highly localized 4$f$ electrons, are good candidates for investigating these interacting charge-, spin- and lattice degrees of freedom. The compound EuPd$_2$Si$_2$ is supposed to be located close to such a  CEP, but slightly on the high-pressure side. Recent studies on Ge-substituted samples \cite{peters2023valence, Wolf2022, Wolf2023} revealed a strong response to hydrostatic pressure, revealing a particular type of CEP. While the valence of Eu in EuPd$_2$Si$_2$ varies by about $\Delta \nu =  0.5$ \cite{mimura2011temperature} when crossing  the phase transition, the lattice parameter  $a$ changes by 2\% \cite{song2023microscopic}, which may cause some damage to the samples \cite{kliemt2022influence}.
\\
With YbInCu$_4$, we identified a Yb-based system where the volume changes at the valence transition are much smaller than in EuPd$_2$Si$_2$ and thus damage to the lattice is likely to be reduced when crossing the transition.
We investigated the ternary intermetallic compound  YbInCu$_4$ which is one of the rare Yb-based cases that exhibits a first-order valence transition at ambient pressure as function of temperature around $T_{\rm V}$ = 42 K \cite{felner1986first}.
In this material, the Yb valence varies by only $\Delta\nu$ = 0.1 \cite{sarrao1996evolution}. Upon cooling, this transition is associated with an increase in lattice volume of 0.5\%, due to the smaller size of the trivalent Yb ion, which is magnetic, compared to the non-magnetic divalent state \cite{sarrao1996evolution}. 
In YbInCu$_4$, the valence state of Yb can be influenced by temperature, pressure \cite{sarrao1998thermodynamics}, magnetic field \cite{katori1994field}, the In-Cu ratio of the samples \cite{loffert1999phase, hiraoka2007neutron} and substitution \cite{sarrao1996evolution}. 
To study possible effects related to critical elasticity in YbInCu$_4$, application of negative chemical pressure is necessary, which shifts the system to the low-pressure side of the CEP, characterized by valence-crossover behavior. For systems close enough to the CEP, the application of He-gas pressure then would allow a detailed investigation of the expected strong-coupling effects. However, the use of chemically-induced pressure via substitution implies the possibility of disorder effects. In addition, there are possible ageing effects due to the volume change at the valence transition.\\
In this work, we used a two-pronged approach: (i) we have grown single crystals of YbIn$_{1-x}$Au$_x$Cu$_4$, a series which was poorly studied so far. (ii) In addition, we have grown Ag-substituted YbInCu$_4$ crystals close to the critical concentration for which the CEP would become accessible at ambient pressure through temperature changes. Subsequently, the crystals in the series (i) and (ii) were characterized by measurements of heat capacity, resistivity, magnetic susceptibility, and ultrasound. These investigations were complemented by experiments under He-gas pressure, enabling a better location of the CEP, and by studies of the effects of disorder and aging.
Previous nuclear magnetic resonance and susceptibility measurements on polycrystalline YbIn$_{1-x}$Au$_x$Cu$_4$ samples with $x = 0.05, 0.1$ revealed that the effect on $T_{\rm V}$ in YbInCu$_4$ was smaller for Au than for Ag substitution \cite{Kojima1992}. 
All parent compounds, YbInCu$_4$, YbAgCu$_4$ and YbAuCu$_4$, crystallize in the cubic AuBe$_5$ (C15b) structure type with the lattice parameters   $a^{\rm In} =  7.158\,$\AA\, \cite{sarrao1996ybin1}, $a^{\rm Ag} =  7.0696\,$\AA\, \cite{severing1990neutron} and $a^{\rm Au} = 7.058\,$\AA\, \cite{casanova1990thermoelectric}.
Furthermore,  YbAuCu$_4$ orders antiferromagnetically below $T_{\rm N} = 0.6\,\rm K$ \enspace \cite{tkeuchi2015themal} ($T_{\rm N} = 0.8\,\rm K$ \enspace \cite{wada2008non}). 
By applying a magnetic field, it is possible to tune the AFM state to a field-induced quantum critical point and further to a non-magnetic Fermi liquid state \cite{tkeuchi2015themal}.
In YbIn$_{1-x}$Ag$_x$Cu$_4$ a CEP can be reached at $T_{\rm crit} = 77\,\rm K$ for $x = 0.195$ \cite{Cornelius1997} with a crossover behavior for $x>0.30$ \cite{sarrao1996evolution} which makes it a candidate system for the study of critical elasticity \cite{Zacharias2015, Gati2016, Wolf2023}.
  \\
In the present work, we have therefore grown and investigated YbIn$_{1-x}$Au$_x$Cu$_4$ samples 
that show a first-order transition (being located at the high-pressure side of the CEP) as well as samples that are in the crossover regime (located at the low-pressure side of the CEP). 
\section{Experiment}\noindent
Single crystals of pure as well as of Ag- or Au-substituted YbInCu$_4$ were grown from In-Cu self-flux as described in Ref.~\cite{sarrao1996ybin1}. The high-purity starting materials Yb (99.99\%, Johnson Matthey), In (99.999\%, Kawecki-Billiton), Cu (99.9999\%, VENTRON GmbH), Au (99.99\%), Ag (9.999\%, Demetron
GmbH) were placed in an aluminum oxide crucible \cite{Canfield2016} sealed in a quartz ampule and heated up to $1100 ^{\circ}$C in a box furnace. After 2~h of  homogenization time, the temperature was reduced to $800 ^{\circ}$C with 2 K/h and the flux was removed subsequently by centrifugation.
The structural analysis of powdered single crystals was performed by powder x-ray diffractometry (PXRD) using a Bruker D8 diffractometer (Bragg-Brentano geometry) with CuK$_{\alpha}$ radiation ($\lambda$ = 1.5406 \AA) at room temperature.
The lattice parameters were refined using the general structure analysis software (GSASII) \cite{toby2013gsas}, and are given in Tab.~\ref{tab:over} and Tab.~\ref{tab:Ag}.\\
Single-crystal diffraction data were collected at 240~K on a STOE IPDS II two-circle diffractometer equipped with the Genix 3D HS microfocus Mo K$\alpha$ source ($\lambda$ = 0.71073 \AA). The crystallographic data and details of the diffraction experiments are given in Tab.~\ref{tab:sc-au} and Tab.~\ref{tab:sc-ag} in Ref.~\cite{supplementalinfo_YbInCu4_2024}. 
We used energy dispersive x-ray spectroscopy (EDX) to characterize the chemical composition of the crystals. We determined the average value from several spots on the same sample. 
The orientation of the samples was determined using a Laue device with white x-rays from a tungsten anode. The crystals usually grow in shape of triangular prisms, representing the (111) planes (Fig.~\ref{fig:Laue} in Ref.~\cite{supplementalinfo_YbInCu4_2024}). The sharp Laue spots indicate a high crystallinity of the samples. 
Differential thermal analysis (DTA) and thermogravimetry (TG) were performed with a Simultaneous Thermal Analysis device (STA 449 C, Netzsch).\\
In this work, the heat capacity was mainly used to determine the nature of the phase transition. Here, the sample experiences a heating pulse in a high vacuum, and the subsequent temperature variation is measured, similar to the conventional semi-adiabatic relaxation method. To determine latent heat, long heat pulses with a typical temperature increase of 8~K were measured and evaluated using the single-lope technique \cite{lashley2003critical}.
Measurements of heat capacity, magnetic susceptibility, and resistivity were performed using the commercial measurement options of a Quantum Design Physical Property Measurement System (PPMS). 
Additional measurements of the magnetic susceptibility
under hydrostatic-pressure conditions were performed using a commercial
superconducting quantum interference device (SQUID)
magnetometer (MPMS; Quantum Design) equipped with
a CuBe pressure cell (Unipress Equipment Division, Institute
of High Pressure Physics, Polish Academy of Science).
The setup covers the temperature range 2 K $\leq T \leq$ 300 K and allows He-gas pressures up to 0.6 GPa. The use of helium as a pressure-transmitting medium ensures truly hydrostatic pressure conditions during the experiment. The pressure cell is connected to a room-temperature He-gas compressor, which serves as a gas reservoir to maintain $p$ = const. conditions upon temperature sweeps. The data have
been corrected for a diamagnetic core contribution, the diamagnetic
contribution of the pressurized He gas and the contribution of the pressure cell. For the ultrasonic experiments, two parallel surfaces perpendicular to the [110] direction were prepared. Two piezoelectric polymer-foil transducers were glued on to these surfaces. To determine the acoustic longitudinal mode, $c_{\rm L}$, sound waves in the frequency range around 75 MHz were propagated along the [110] direction. By using a phase-sensitive detection technique as described in Ref.~\cite{luthi1994electron} the relative change of the sound velocity were measured as the function of temperature.

\section{Crystal Growth}
\begin{figure}[htbp]
    \centering
    \includegraphics[width=0.5\textwidth]{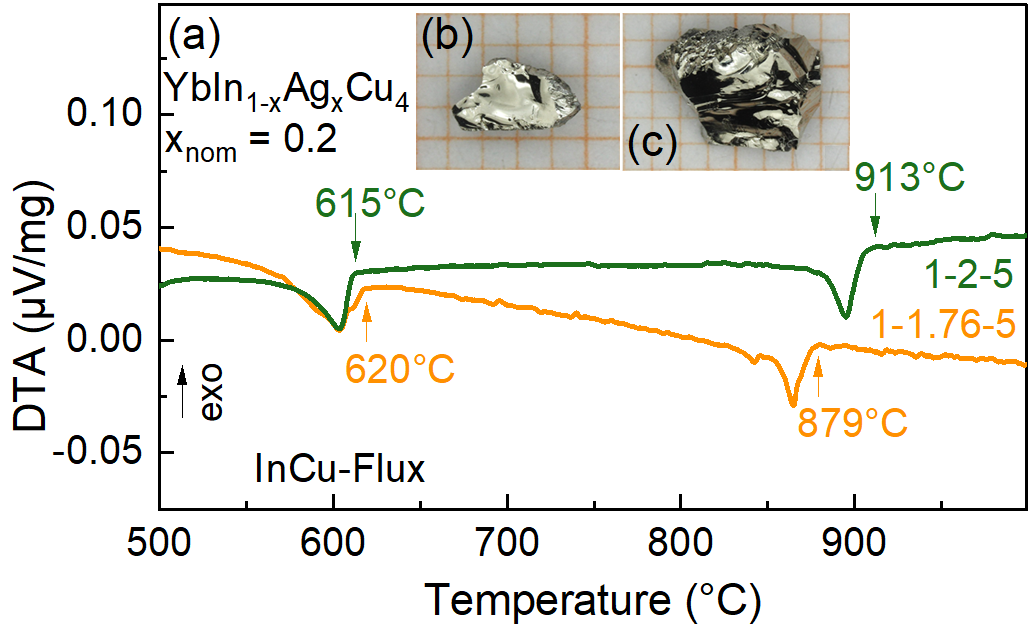}
    \caption{YbIn$_{1-x}$Ag$_x$Cu$_4$, $x_{\rm nom}$ = 0.20: (a) Differential thermal analysis signal of the second cooling for two initial compositions. The green data correspond to the 1-2-5 while the orange data represent the 1-1.76-5 melt composition. Depending on the composition, the compound solidifies at around 900~$^\circ$C while the flux solidifies at lower temperatures. The inset shows typical samples from two growth experiments with $x_{\rm nom}$ = 0.20: (b) for an initial composition of 1-2-5 and (c) for an initial composition of 1-1.76-5.}
    \label{fig:DTA}
\end{figure}

\noindent
We analyzed the crystal growth with regard to its solidification temperature using differential thermal analysis (DTA).
For this comparative analysis 1g of the elements was weighed in using various stoichiometries and sealed  under vacuum in a quartz ampule to prevent evaporation of Yb which is highly volatile at elevated  temperatures. The ampule was heated up to $1100 ^\circ$C with a heating/cooling rate of 10 K/min.
We compared two In-Cu fluxes with the different initial compositions of the melt, Yb:In:Cu = 1:2:5 (1-2-5) and Yb:In:Cu = 1:1.76:5 (1-1.76-5).  The cooling curves of the second temperature cycle of samples with an Ag concentration of x$_{\rm nom}$ = 0.20 were analyzed as shown in Fig.~\ref{fig:DTA}.  
The DTA signal contains mainly two features:
The first dip is assigned to the solidification of the Ag-substituted compound YbInCu$_4$ and occurs, depending on the initial composition, at $913 ^{\circ}$C for the 1-2-5 or at lower temperatures of $879 ^{\circ}$C for the 1-1.76-5 melt composition. The second dip at $615 - 620 ^{\circ}$C is related to the solidification of the In-Cu flux \cite{liu2002thermodynamic}.
Compared to the growth from a stoichiometric composition with $T_{\rm sol} = 970 ^{\circ}$C \cite{fischbach1998thermoanalytical}, the Ag-substituted target compound solidifies at lower temperatures in the In-Cu flux. Using less In, the solidification temperature seems to decrease even further. 
A similar behavior was also observed in Ref.  \cite{fischbach1998thermoanalytical} under variation of the Yb content. \\
Furthermore, we see in our growth experiments that the 1-1.76-5 composition produces supposedly larger crystals. Two typical examples are shown in the inset of Fig.~\ref{fig:DTA}. This suggests that using less In has a positive effect on the yield of the growth experiments. A unique feature to this observation are the Ag-substituted samples with $x_{\rm nom} = 0.136$. Here, instead of several crystals, one large crystal was formed. This crystal will be analyzed in more detail below. \\

\section{Results and Discussion}\noindent
\subsection{YbIn$_{1-x}$Au$_x$Cu$_4$} \noindent
\subsubsection{Structural and chemical analysis}\noindent
All Au-substituted samples were grown using the initial melt composition Yb-(In+Au)-Cu = 1-1.76-5. For the characterization of unsubstituted samples with this composition, see Sec.~\ref{sec:YbInCu4-HC} and Fig.~\ref{fig:HC_unsub} in the supplemental material \cite{supplementalinfo_YbInCu4_2024}.
Here, we show the analysis of samples with 5 different Au-substitution levels up to $x_{\rm nom}$ = 0.30. The results are summarized in Tab.~\ref{tab:over}.
The nominal Au concentration $x_{\rm nom}$ was compared to the concentration $x_{\rm EDX}$ measured by EDX which was determined with $x_{\rm EDX}$ = Au/(Au+In). The error of $x_{\rm EDX}$ was determined only from the statistics of several measurements.
Our analysis revealed that the Yb content remains constant for the samples, which is consistent with previous work \cite{loffert1999phase, moriyoshi2003crystal}.
It was found that about 1.5 times of the initial amount of Au in the melt is built into the crystal.
The single-crystal (SC) XRD data confirmed the cubic $F\overline{4}3m$ structure and revealed that between two Au-substituted samples with $x_{\rm nom}$ = 0.10 and $x_{\rm nom}$ = 0.30 the lattice parameter differs significantly, by 0.32\% (Tab.~\ref{tab:sc-au} in Ref.~\cite{supplementalinfo_YbInCu4_2024}). First, we assumed that Au partly substitutes In in its  position, as it is supposed for Ag-substitution in Ref. \cite{sarrao1996evolution}. Then we found two alternative ways of further improvement of the refinement residuals. 
In the model A we assumed the In/Au mixture in the position of indium and Cu/Au mixture in the position of copper (Yb(In$_{1-x}$Au$_x$)(Cu$_{1-y}$Au$_y$)$_4$). In turn, in the model B we also assumed the In/Au mixture in the position of indium, but, alternatively, the Cu/In mixture in the position of copper (Yb(In$_{1-x}$Au$_x$)(Cu$_{1-y}$In$_y$)$_4$). Both models end up with significantly different chemical composition but with very close residual values (Tab.~\ref{tab:sc-au} in Ref.~\cite{supplementalinfo_YbInCu4_2024}). Therefore, from the SC-XRD data we cannot judge, what is the additional atom in the position of Cu, is it Au, In or both. We found that the Yb site is fully occupied for both samples.  
Our powder diffraction data reveals that the lattice parameter $a$ remains constant up to a substitution level of $x_{\rm nom}$ = 0.15 as shown in Tab.~\ref{tab:over}.  At $x_{\rm nom}$ = 0.30 the lattice parameter appears to decrease. A similar  behavior was already reported in the series with Ag substitution \cite{sarrao1996ybin1}. All these results indicate structural disorder in the crystals.
\begin{table}[]
    \centering
    \begin{tabular}{|c|c|c|c|c|}
        \hline
         $x_{\rm nom}$&$x_{\rm EDX}$& $a$   &$T_{\rm V}$  &$T^{\prime}_V$         \\
         &&  [\AA] & [K] & [K]
         \\
         \hline\hline
         0& 0 & 
         7.155(5)& 43&
         \\
         0.05& 0.099(5) &
         7.158(7)& 50&
         \\
         0.08& 0.132(3)&
         7.159(2)&64& 
         \\
         0.10& 0.153(5) & 
         7.155(5)& &75
         \\
         0.15&0.236(3) &
         7.160(5)&& 84
         \\
         0.30&0.372(3) &
         7.150(5)&&- 
         \\
         \hline
    \end{tabular}
    \caption{Overview of the nominal, $x_{\rm nom}$, and measured, $x_{\rm EDX}$, Au concentration of the compound YbIn$_{1-x}$Au$_x$Cu$_4$ for different substitution levels $x$ together with the corresponding lattice parameters $a$ as determined from PXRD, the transition temperature $T_{\rm V}$, and  the crossover temperature $T^{\prime}_V$.}
    \label{tab:over}
\end{table}

\subsubsection{Heat capacity }
\begin{figure}
    \centering
    \includegraphics[width=0.5\textwidth]{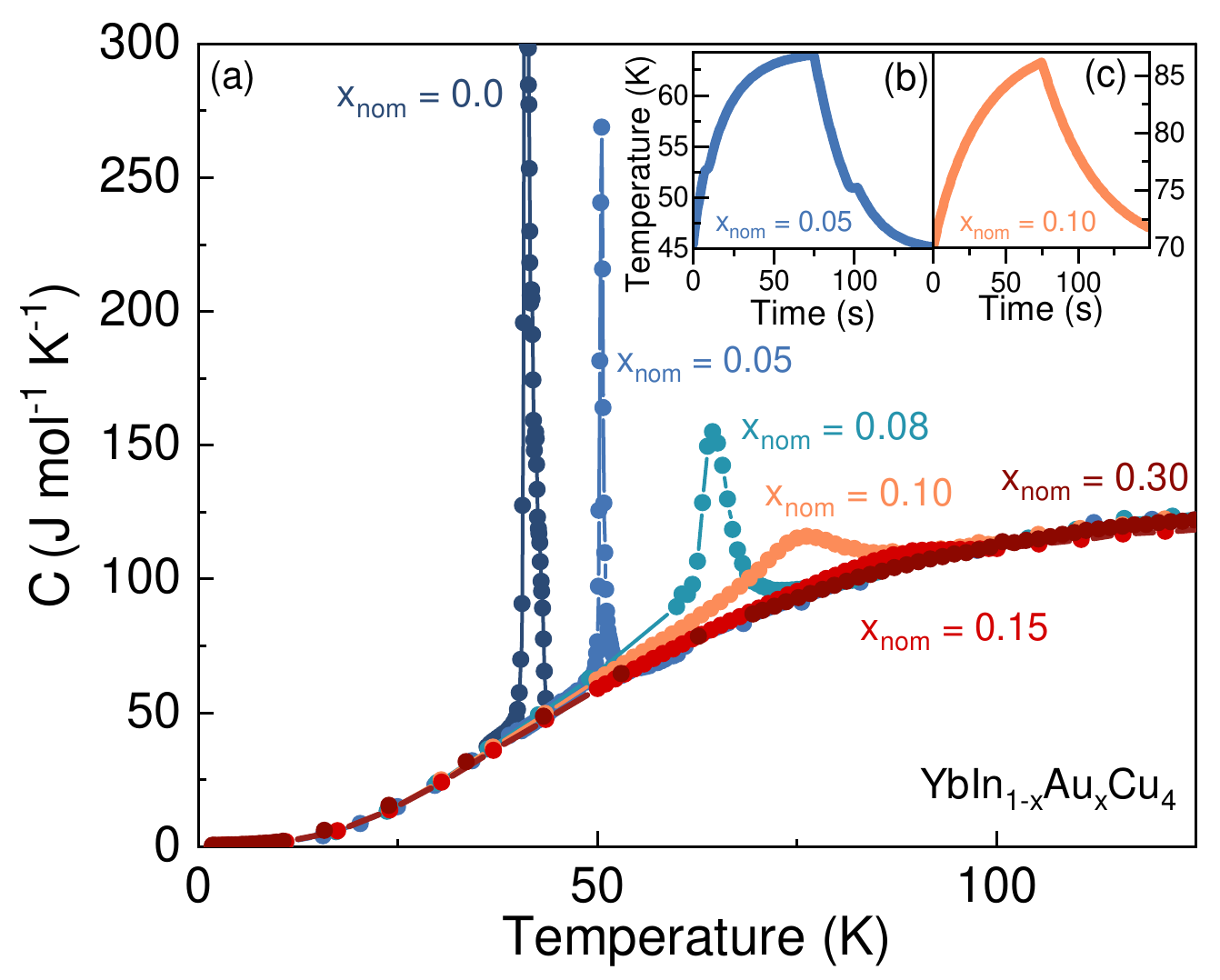}
    \caption{YbIn$_{1-x}$Au$_x$Cu$_4$: (a) Heat capacity as a function of temperature for different Au concentrations. (b) Horizontal parts, corresponding to latent heat, indicates a first-order transition measured on a sample with $x_{\rm nom}$ = 0.05. (c)  The continuous response to the heat pulse is associated with a valence crossover. }
    \label{fig:Heat}
\end{figure}

\noindent
The heat capacity of YbIn$_{1-x}$Au$_x$Cu$_4$ is shown in Fig.~\ref{fig:Heat}(a) for samples with different substitution levels. The unsubstituted sample displays a sharp peak at 41 K which is a characteristic of a first-order phase transition.
With increasing Au concentration, the transition temperature increases and the peak broadens above $x_{\rm nom}$ = 0.05 until it completely vanishes for $x_{\rm nom} = 0.3$. A first-order phase transition is associated with the occurrence of latent heat at the phase transition, which is visible as a kink in the heat-pulse signal (see insets of Fig.~\ref{fig:Heat}). 
Our data show a first-order phase transition for samples up to a substitution level of $x_{\rm nom}$ = 0.08 (blue symbols). The heat pulse of the sample with the first-order transition with $x_{\rm nom}$ = 0.05 is shown in Fig.~\ref{fig:Heat}(b). 
We observe a valence crossover in the sample with $x_{\rm nom}$ = 0.10 (orange symbols) which exhibits a continuous response to the heat pulse as shown in Fig.~\ref{fig:Heat}(c). 
As the Au concentration increases, the heat-capacity anomaly becomes progressively less significant (red symbols). When the Au concentration reaches $x_{\rm nom}$ = 0.30 (dark red symbols) the 
peak is no longer observed. 

\begin{figure}
    \centering
    \includegraphics[width=0.5\textwidth]{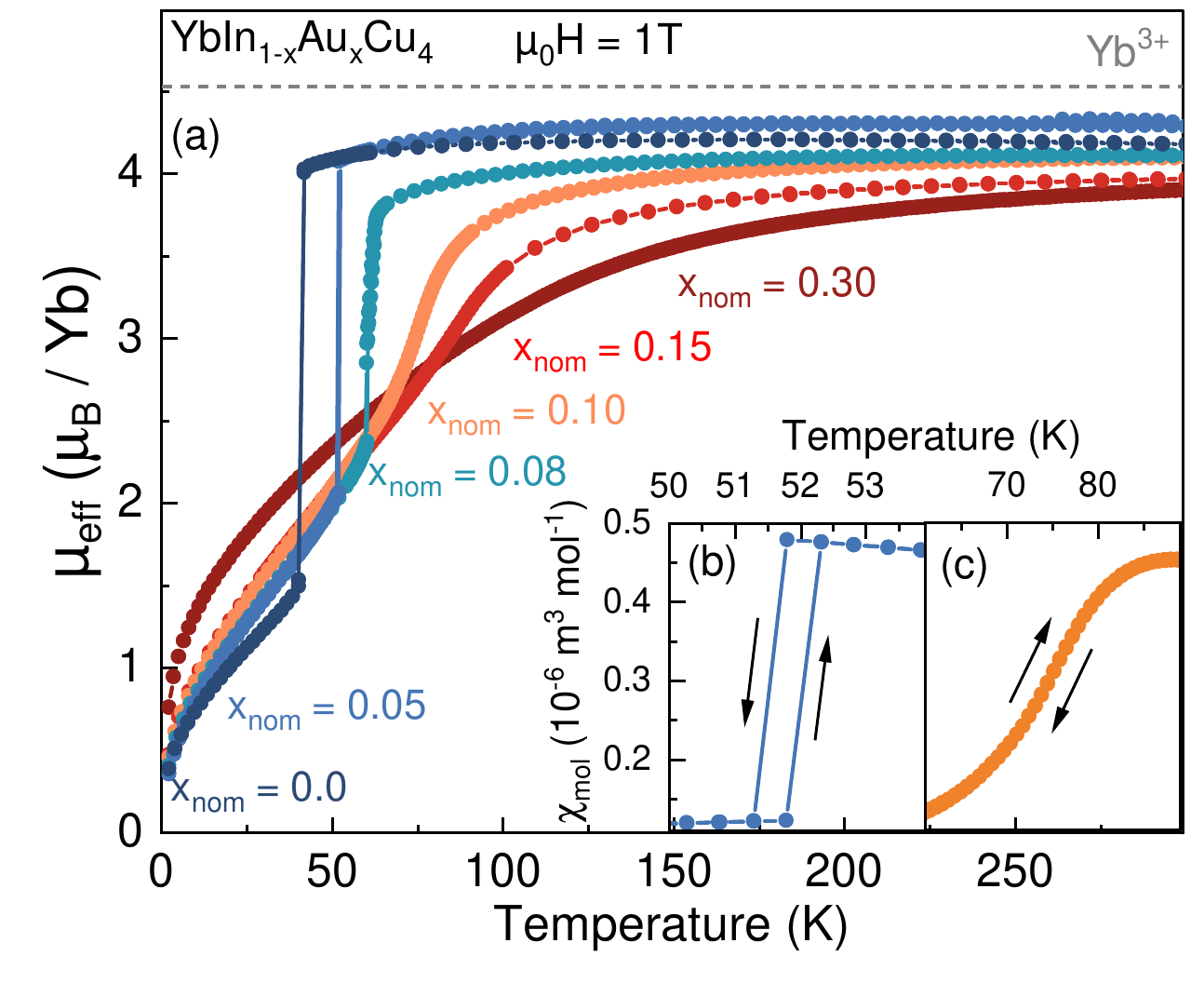}
    \caption{YbIn$_{1-x}$Au$_x$Cu$_4$: (a) Effective magnetic moment as a function of temperature for different substitution levels extracted from the data shown in Fig.~\ref{fig:sus} in Ref.~\cite{supplementalinfo_YbInCu4_2024}. (b) Hysteresis indicating a first-order transition recorded for $x_{\rm nom} = 0.05$. (c) The $x_{\rm nom} = 0.10$ sample does not show a hysteresis, which indicates a valence crossover.}
    \label{fig:Au_mB_Hys}
\end{figure}
\subsubsection{Magnetic susceptibility}\noindent
The magnetic susceptibility as a function of temperature was measured in a field of $\mu_0H = 1\,\rm T$. From this data we calculated the effective magnetic moment as a function of temperature, Fig.~\ref{fig:Au_mB_Hys}(a), according to $\mu_{\rm eff} = \sqrt{{3 k_{\rm B} T \chi_{\rm mol}}/{N_{\rm A} \mu_0 \mu_{\rm B}^2}},$ 
with the Boltzmann constant $k_B$, the Avogadro number N$_{\rm A}$, the vacuum permeability $\mu_0$ and the Bohr magneton $\mu_B$ from data presented in Fig.~\ref{fig:sus} in Ref.~\cite{supplementalinfo_YbInCu4_2024}.
At high temperatures, $\mu_{\rm eff}$ remains approximately constant following a Curie-Weiss law and undergoes a sharp drop at the phase transition when lowering the temperature. For temperatures $T<T_{\rm V}$,  $\mu_{\rm eff}$ drops further. Consistently with the heat capacity, the valence transition shifts to higher temperatures with increasing Au-substitution level and the drop at the phase transition becomes broader. The unsubstituted sample shows $\mu_{\rm eff}$ = 4.18\,$\mu_{\rm B}$ at high $T$, which is close to that of the trivalent Yb state. With higher Au concentration, the absolute value of $\mu_{\rm eff}$ decreases. Consequently, the sample with the highest substitution level shows the lowest effective moment of order 4.0~$\mu_B$.   
Deviations from the Curie-Weiss law at high temperatures can be attributed to non magnetic contributions in the sample or an inaccurate molar mass
determination due to flux inclusions. 
At the first-order valence transition, a hysteresis between cooling and heating curves of $\approx 1 \,\rm K$ is observed as it can be seen in Fig.~\ref{fig:Au_mB_Hys}(b) for a sample with $x_{\rm nom} = 0.05$. In contrast, no hysteresis is observed for samples in the crossover region, Fig.~\ref{fig:Au_mB_Hys}(c). 
It should be noted that the cooling rate during the measurement influences the size of the hysteresis. Thus, also data recorded at samples in the crossover regime might show a hysteresis in case the temperature is swept too fast ($\geq$3\,K/min). \\
\begin{figure}[htpb]
    \centering
    \includegraphics[width = 0.5\textwidth]{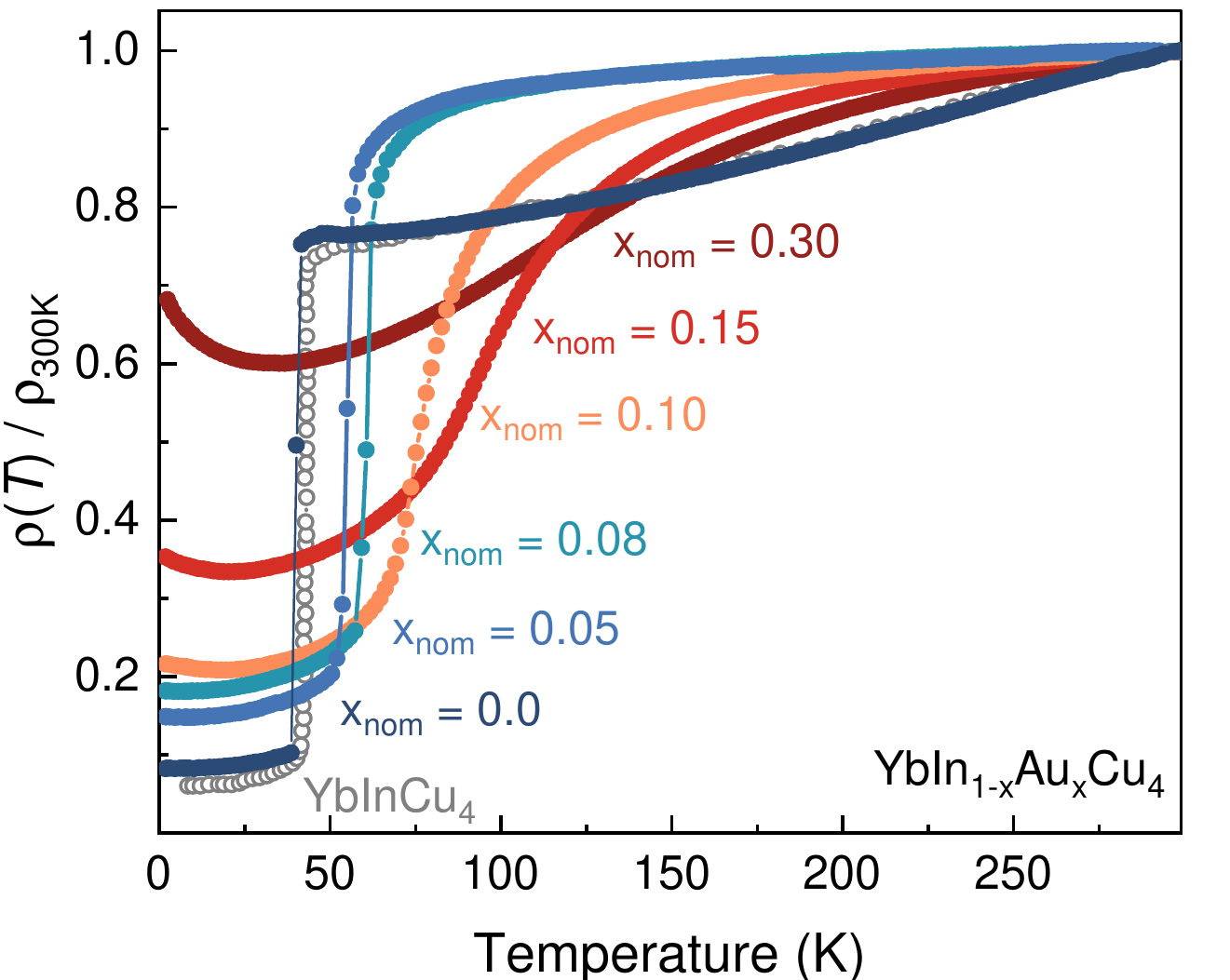}
    \caption{YbIn$_{1-x}$Au$_x$Cu$_4$: Normalized electrical resistivity as a function of temperature for samples with different Au-substitution levels. The data of pure YbInCu$_4$ (open gray symbols) \cite{sarrao1996evolution} are shown for comparison.}
    \label{fig:res}
\end{figure}
\subsubsection{Resistivity}\noindent
The electrical resistivity of the Au-substituted samples, Fig.~\ref{fig:res}, was studied between 1.8\,K and 300\,K. 
The room temperature resistivity determined from the first cooling curve for samples with $x_{\rm nom}$ = 0.05 is $\rho(300\,\rm K) = 150\, \mu\Omega$cm which agrees well with that of the unsubstituted samples with $\rho(300\,\rm K) = 150\,\mu \Omega$cm \cite{sarrao1996evolution} and increases slightly for $x_{\rm nom}$ = 0.30 to $\rho(300\,\rm K) = 177\,\mu\Omega$cm.
For high temperatures $300\,{\rm K}>T>T_{\rm V}$, the electrical resistivity shows metallic behavior and displays a sudden decrease at the transition temperature. \\
Due to the volume change caused by the transition to the lower-valence state of Yb, the absolute value of the electrical resistivity increases with each passing of the phase transition, possibly due to internal strain as described in Ref.~\cite{sarrao1996evolution}. Therefore, for better comparability of samples with different Au concentrations, the respective first cooling curves are shown here. Upon approaching the critical concentration, the Yb volume change becomes smaller and smaller, leading to a reduction of the intrinsic strain in the sample when crossing the transition.
The electrical resistivity of samples in the crossover region shows no signatures of damage when passing through the crossover.  
The residual resistivity $\rho_0$  increases with increasing substitution level, indicating that the disorder in the crystals also increases. For $x_{\rm nom}$ = 0.10, RRR$_{\rm 2K}=\rho_{300\rm K}/\rho_{2\rm K}$ corresponds to about 5, which is comparable to the value of 6.5 for $x$ = 0.27 a sample of the crossover region in the Ag-substitution series. Therefore, the disorder is comparable among the two series.
\\
In accordance with heat capacity and magnetization results, the phase transition or crossover temperature shifts to higher temperatures as the Au concentration increases. The sample with $x = 0.3$ does not show any anomaly down to 1.8\,K. 

\subsection{YbIn$_{1-x}$Ag$_x$Cu$_4$}\noindent
\subsubsection{Structural  analysis}\noindent
We used mainly two initial melt compositions for Ag-substituted growth experiments, with the results of the 1-2-5 being similar to that published in Ref.~\cite{sarrao1996evolution}. Furthermore, we reduced the indium content and used the initial composition of the melt 1-1.76-5. A summary of all Ag-substituted samples is shown in Tab.~\ref{tab:Ag}. The integrated Ag content $x_{\rm EDX}$ of all samples is approximately half the nominal value $x_{\rm nom}$. The cubic $F\overline{4}3m$ structure was verified for all samples using PXRD. The lattice parameter $a$ is shown in Tab.~\ref{tab:Ag}. We found that this parameter remains approximately constant in the investigated composition range, in accordance with Ref. \cite{sarrao1996ybin1}.   
\\
\begin{table}[]
    \centering
    \begin{tabular}{|c|c|c|c|c|c|}
    \hline
        initial composition & $x_{\rm nom}$& $x_{\rm EDX}$ &$ a$ & $T_{\rm V}$  &$T^{\prime}_V$  \\
         & &  &[\AA]& [K] & [K] \\
        \hline
        \hline
        1-2-5 & 0& 0 &
        7.155(4)&42& \\
        1-2-5 & 0.10&0.057(3)  & 
        7.156(3) &55& \\
        1-2-5 & 0.20& 0.111(3) & 
        7.155(4)&80 &\\
        1-2-5 & 0.25& 0.14(1) &
        7.154(5)&87 &\\
        1-2-5 & 0.27&0.173(7)  &
        7.158(3)&&98 \\
        \hline
        1-1.76-5 & 0&0 &
        7.155(3)& 41&\\
        1-1.76-5 & 0.136& 0.088(4) & 
        7.159(6)& 64-70&\\
        1-1.76-5 & 0.20& 0.122(4) & 
        7.153(5)& &84\\
        \hline
        1-1.5-5 & 0.20&0.132(3)  & 7.150(2)
        & &91.4\\
        \hline
        1-1.76-5.5 & 0.20& 0.118(3) & 7.149(4)
        &82.5 & \\
        \hline
    \end{tabular}
    \caption{Summary of the nominal ($x_{\rm nom}$) and measured ($x_{\rm EDX}$) Ag concentration of the compound YbIn$_{1-x}$Ag$_x$Cu$_4$ grown from different initial compositions of the melt together with the lattice parameter $a$ as determined from PXRD, the transition temperature $T_{\rm V}$, and the crossover temperature $T^{\prime}_V$. The transition temperatures were extracted from Fig.~\ref{fig:HC_Agseries} and Fig.~\ref{fig:HC_ag}. For the sample with $x_{\rm nom}$ = 0.136, the transition temperature varies over the large crystal due to a non-homogeneous Ag distribution. }
    \label{tab:Ag}
\end{table}
The single crystal analysis reveals that the value of the lattice parameter for two Ag-substituted crystals differs only very slightly, within 0.06\%, which can be explained by the accuracy of 2-cycles single-crystal diffractometer (Tab. ~\ref{tab:sc-ag} in Ref.~\cite{supplementalinfo_YbInCu4_2024}). The refinement of the site occupancy factors in both samples with $x_{\rm nom}$ = 0 and $x_{\rm nom}$ = 0.20 shows no signs of nonstoichiometry. The attempted refinement of the Cu/Ag and Yb/Ag mixed site compositions in the $x_{\rm nom}$ = 0.20 sample ends up with 100\% of Cu and Yb, respectively, proving that Ag substitutes neither Cu nor Yb. The refinement of In/Ag mixture in the position of In fails due the too small difference of the atomic scattering factors of these elements. 
Both methodes PXRD and the SC-XRD have different systematic errors, including different geometries and refinement methodes, so that there is a difference in the lattice parameter $a$. 
Additionally, the SC analysis measures a selected small $\approx 10~\mu$m sample with presumably one sharp concentration, while the PXRD analysis measures an average value over many concentrations. 
Note that we also found an indium excess in the Ag-substituted samples in our EDX measurements. Furthermore, the (In+Ag)/Cu ratio varies between the different samples within one batch, as shown in Fig.~\ref{fig:tenaerag}. 
However, this may also be related to the accuracy of the EDX measurements. \\
\subsubsection{Heat capacity} 
\begin{figure*}
    \centering
    \includegraphics[width=0.47\textwidth]{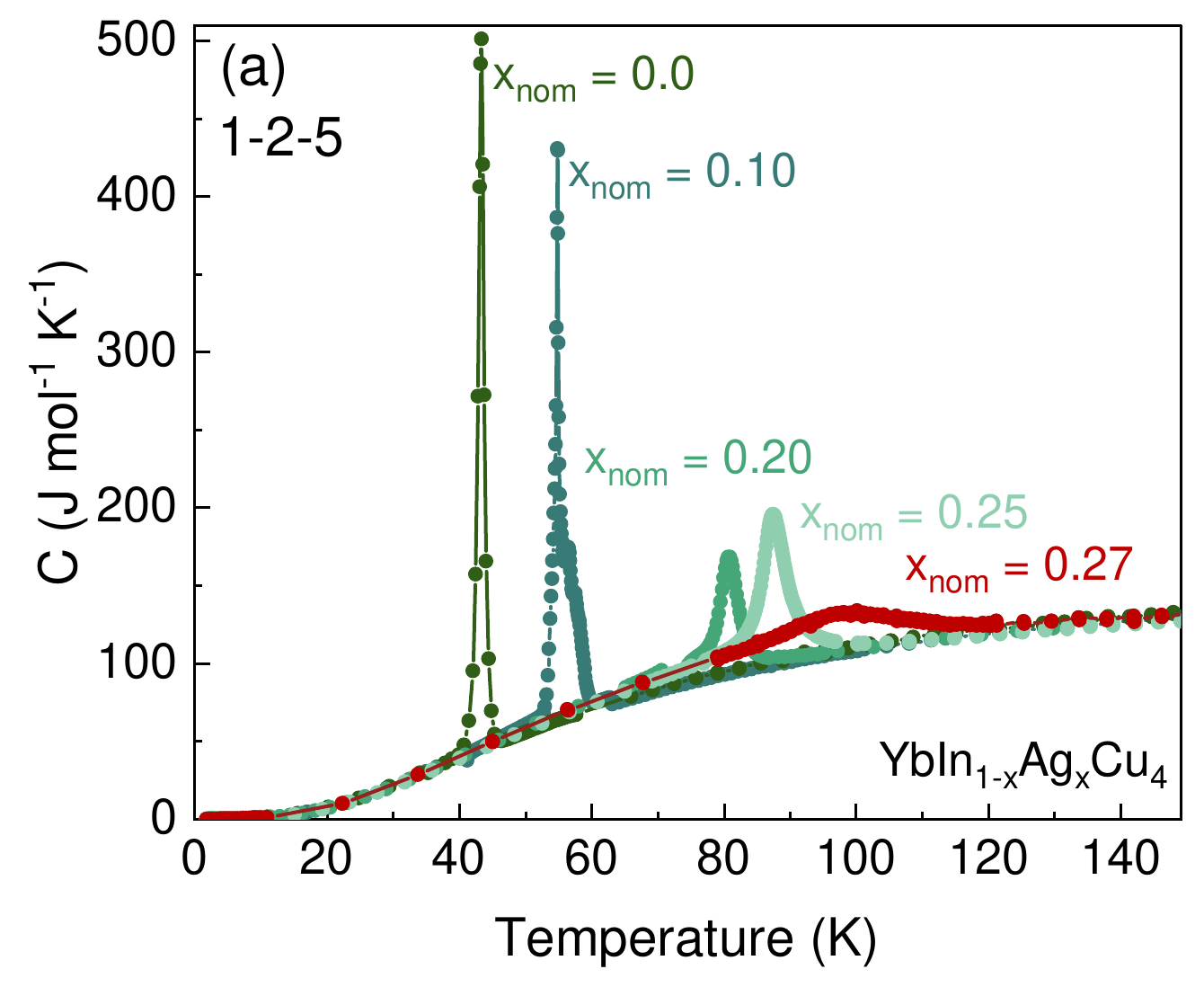}
    \includegraphics[width=0.48\textwidth]{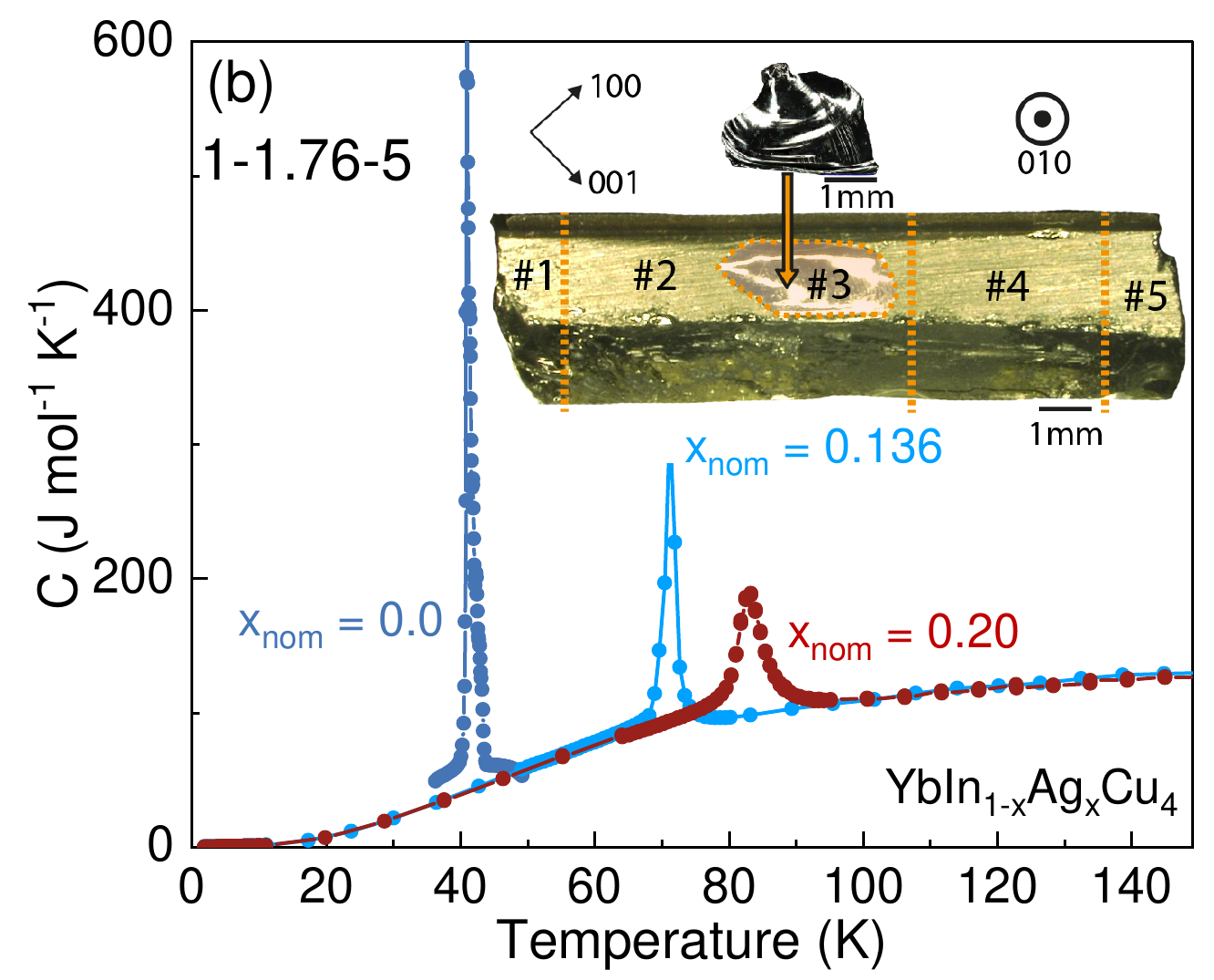}
    \caption{Heat capacity measured on YbIn$_{1-x}$Ag$_{x}$Cu$_4$ samples grown from a melt with the initial composition (a) 1-2-5 and (b) 1-1.76-5. Blue and green  symbols indicate data where latent heat was detected in the heat capacity, red symbols indicate data taken on samples in the valence-crossover regime. The inset in (b) shows a part of the large sample with $x_{\rm nom}$ = 0.136 grown from the 1-1.76-5 composition.}
    \label{fig:HC_Agseries}
\end{figure*}

\noindent
\label{chap:HC_Ag}\noindent
We performed several experiments investigating samples with different Ag-substitution levels  grown from different initial melt compositions. An overview of the performed growth experiments is given in Tab.~\ref{tab:Ag}. A first substitution series using the composition 1-2-5 was reported in literature \cite{sarrao1996evolution} and the results of these previous experiments were reproduced successfully here. The heat capacity as a function of temperature is shown in Fig.~\ref{fig:HC_Agseries}(a).
We observe that the peak in the heat capacity of the unsubstituted sample  shifts to higher temperatures and becomes broader at higher Ag concentrations. A first-order phase transition (identified by the observation of latent heat) can be found up to an Ag concentration of $x_{\rm nom}$ = 0.25 (green) yielding a transition temperature of $T_{\rm V}$ = 87\,K. A sample with $x_{\rm nom}$ = 0.27 displays a valence crossover at $T^{\prime}_V$ = 98\,K (red). Further experiments on unsubstituted samples show that 1-1.76-5 samples present a supposedly sharper transition, while the transition temperature is comparable (Fig.~\ref{fig:HC_unsub} in Ref.~\cite{supplementalinfo_YbInCu4_2024}). With increasing Ag concentration, the peaks shift to higher temperatures.  However, for 1-1.76-5 samples a valence crossover is achieved at a lower temperature $T^{\prime}_V$ = 84\,K (red) for a lower Ag concentration of $x_{\rm nom} = 0.2$. The sample with $x_{\rm nom}$ = 0.136, inset of Fig.~\ref{fig:HC_Agseries}(b), was taken from an unusual large crystal shown in Fig.~\ref{fig:Laue} in Ref.~\cite{supplementalinfo_YbInCu4_2024}. The samples cut from this crystal exhibit different transition temperatures, which leads us to the conclusion that different Ag concentrations are present in one sample. 
\subsubsection{Magnetic and elastic properties of the valence transition close to the CEP} \noindent
\label{chap:valencetransition}\noindent
A large ($\approx$ 12~\,mm) sample with $x_{\rm nom}$ = 0.136 (cut from the sample shown in \ref{fig:Laue}), shown in the inset of Fig.~\ref{fig:HC_Agseries}(b), was cut into several pieces along the [110] direction for ultrasonic studies. We found that different samples exhibit a range of several transition temperatures.
The pieces \#2 (data not shown) and \#4 
were used for ultrasonic experiments, whereas the pieces \#1,\#3 and \#5 were used to determine their respective magnetic properties via SQUID measurements.
\begin{figure}
    \centering
    \includegraphics[width=1\linewidth]{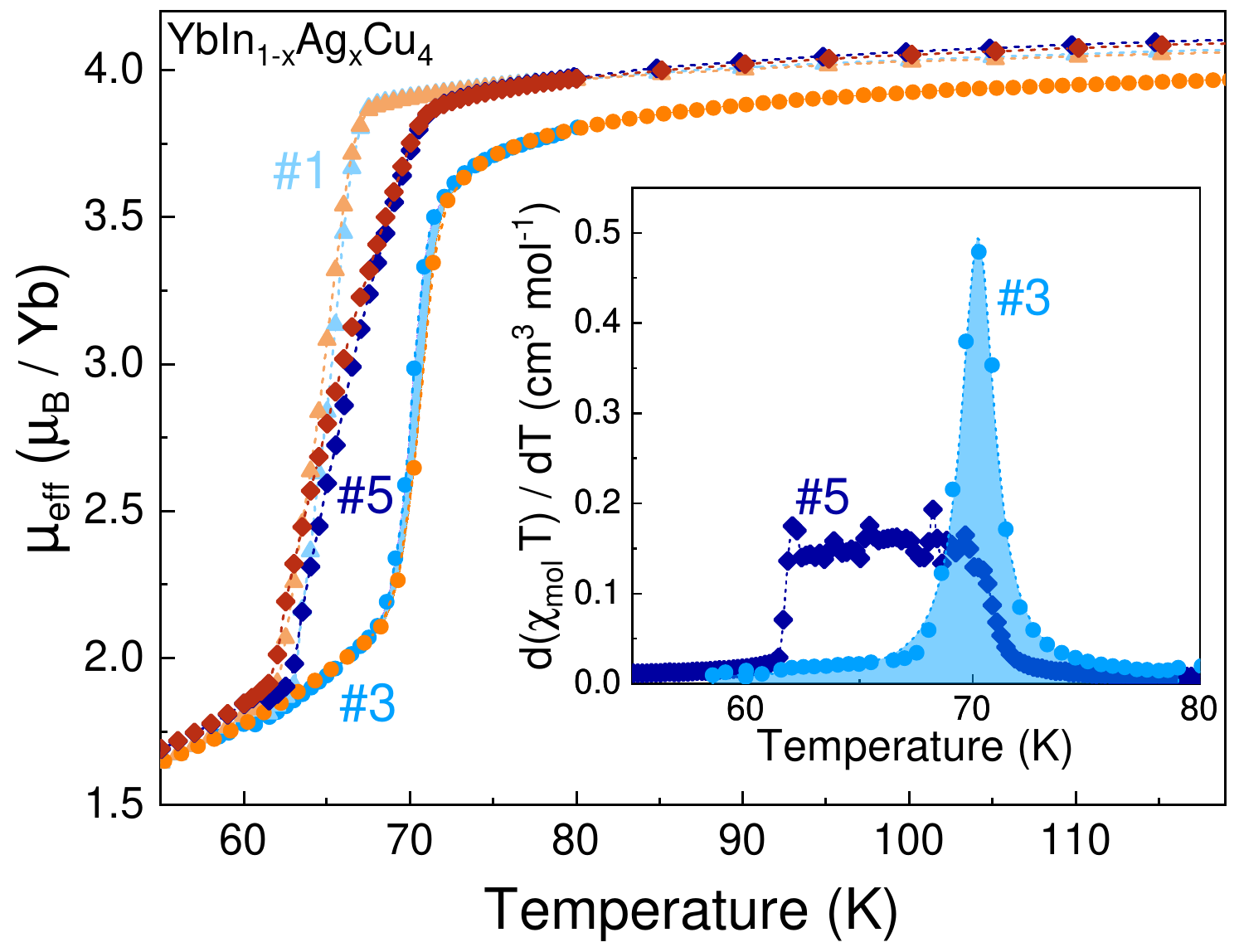}
    \caption{Effective magnetic moment $\mu_{\rm eff}$ of different samples (\#1 
    triangles, \#5 diamonds, and \#3 circles) cut from the large single crystal  with $x_{\rm nom}$ = 0.136 (see inset of Fig.~\ref{fig:HC_Agseries}(b) for the labeling of the individual samples) as a function of temperature. The data were taken at ambient pressure during cooling (blue symbols) and with increasing temperature (red/orange symbols).  Inset: Temperature derivative of the quantity d($\chi_{\rm mol}T$)/d$T$ as a function of temperature. The two data sets were taken at ambient pressure during cooling of sample \#5 (dark blue diamond) and sample \#3 (blue circles). The broken line enclosing the blue-shaded area represents a Lorenzian fit to the experimental data.}
    \label{fig:neff}
\end{figure}\noindent
The main panel of Fig.~\ref{fig:neff} shows the temperature dependence of the effective magnetic moment $\mu_{\rm eff}$ in the temperature range $55\,{\rm K} \leq T \leq 120\,\rm K$ for the samples \#1, \#3 and \#5. 
Around 120 K and above, the effective magnetic moment is nearly temperature independent for all single crystals investigated. Such a behavior was already observed in Ref.~\cite{sarrao1996evolution} on single crystals with different Ag-substitution levels, yielding a $\mu_{\rm eff}$ at these high temperatures, which is reduced with increasing Ag concentration. Based on the systematics derived in Ref.~\cite{sarrao1996evolution}, we were able to give an estimate of the differences in the effective Ag concentration $x_{\rm eff}$ of the individual samples by using their $\mu_{\rm eff}$ values at high temperatures. The samples \#1 (orange/blue triangles in Fig.~\ref{fig:neff}) and \#5 (red/blue diamonds in Fig.~\ref{fig:neff}), i.e., the bulky outer pieces, exhibit the same $\mu_{\rm eff}$ values at 120 K, reflecting a nearly identical Ag concentration. In contrast, the sample \#3, a small plate-like piece from the (100)-surface in the middle of the main sample, shows a reduced $\mu_{\rm eff}$ value and thus a higher Ag concentration (orange/blue circles). Note that this sample \#3 was also used for the heat-capacity experiments, revealing clear evidence for a first-order valence transition. For this very sample \#3, we determine a transition temperature of $T_{\rm V}$ = 70.2~K, corresponding to the position of the maximum in d($\chi_{mol}T$)/d$T$, displayed in the inset of Fig.~\ref{fig:neff}. The width of the valence transition, quantified by taking the full width at half maximum of a Lorentz curve fitted to the data, amounts to 1.8 K. Note that this value is remarkably small for the Ag-substituted single crystal \#3. \\
Despite the first-order transition for crystal \#3, as determined by heat capacity, we recognize some rounding in $\mu_{\rm eff}$($T$) above and below the valence transition, and a lack of a thermal hysteresis on the temperature scale in Fig.~\ref{fig:neff}. The effective magnetic moment of the samples \#1 and \#5 exhibits a somewhat different temperature dependence around the valence transition. The first-order character of the valence transition is reflected in sharp kinks at the on- and offsets for both crystals. In addition, and at variance to the behavior seen for sample \#3 in Fig.~\ref{fig:neff}, the change in $\mu_{\rm eff}$($T$) for samples \#1 and \#5 occurs over significantly wider temperature ranges. Like for \#3, a thermal hysteresis cannot be resolved on the temperature scale of the main panel of Fig.~\ref{fig:neff}. The temperature range in which the valence transition takes place in sample \#5 can be determined by considering the quantity d$(\chi_{mol}T)/$d$T$. As shown in the inset of Fig.~\ref{fig:neff}, based on the onset of the anomaly, the transition range extends significantly from (72.6 $\pm$ 0.1)~K to (61.0 $\pm$ 0.1)~K. Furthermore, it is obvious that these data cannot be fitted using a single Lorenzian. In contrast, the data seem to represent an array of individual Lorentz-lines, corresponding to a distribution of different Ag concentrations. 
Based on these observations, we conclude that for the single crystal with $x_{\rm nom }$ = 0.136  within its average Ag concentration of $x_{\rm EDX}$ = 0.088 there is a slight increase in $x_{\rm eff}$ between the core and the surface of the sample, reflected in an overall somewhat reduced transition temperature. As a consequence of the inhomogeneity in $x_{\rm eff}$, the width of the transition for the rather large chunks \#1 and \#5 is also enhanced. In contrast, the small fragment \#3 appears to have a rather homogeneous Ag distribution, reflected in a very narrow transition. The properties of the sample can be attributed to the growth method. As growing takes place in the flux, the composition of the melt changes with the time of growth and the size of the crystals. According to our EDX analysis, only half of the available Ag is incorporated in the samples, leading to an enrichment of Ag in the melt. Correspondingly, we obtain a gradient of Ag concentrations especially for the large single crystals like the one displayed in the inset of Fig.~\ref{fig:HC_Agseries}(b) with the core having a lower Ag content leading to $T_{\rm V}\approx60\,\rm K$ and an outer shell with higher Ag content leading to $T_{\rm V}\approx70\,\rm K$. 
\begin{figure}
    \centering
    \includegraphics[width=1\linewidth]{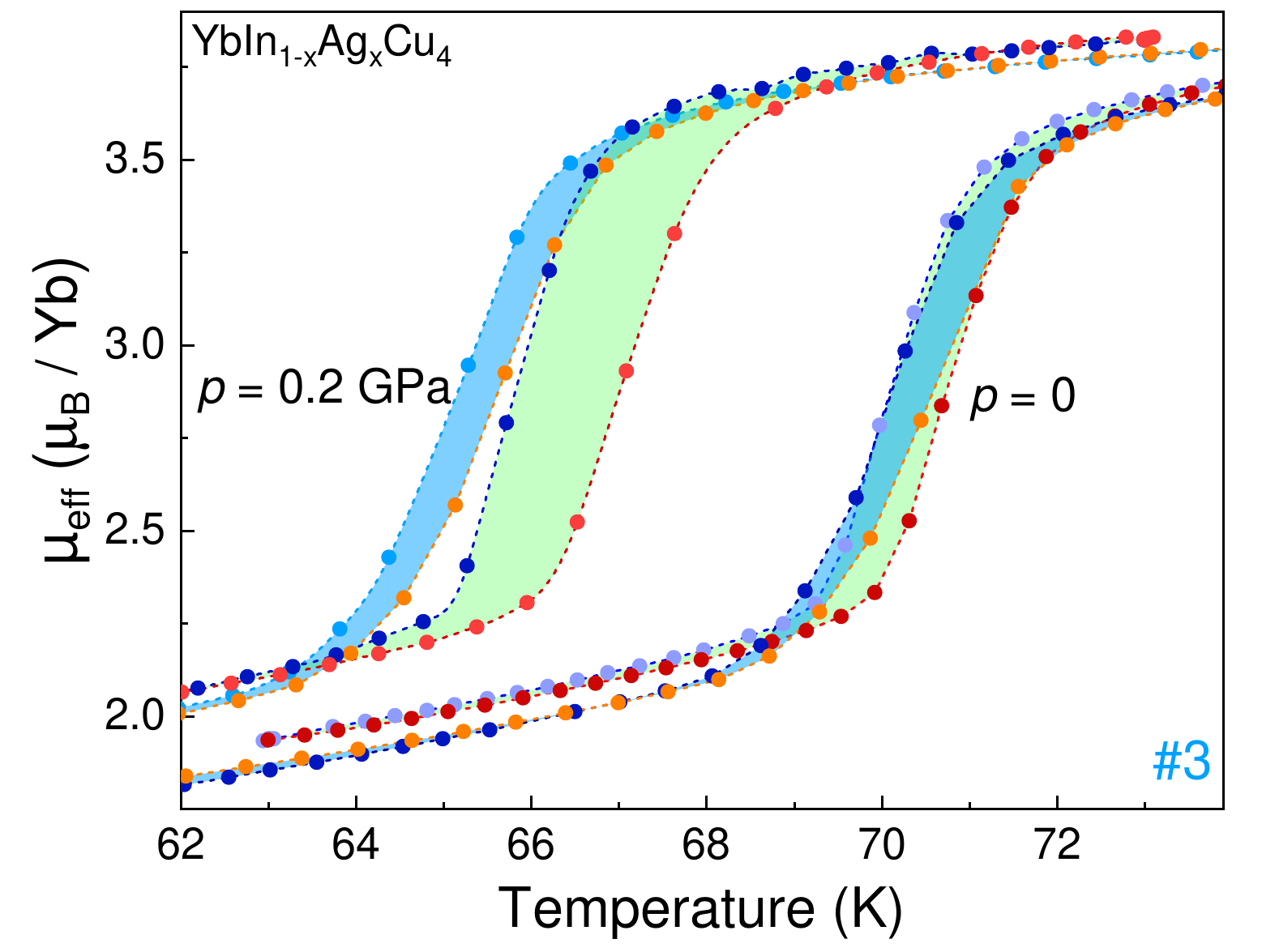}
    \caption{Enlarged representation of $\mu_{\rm eff}$ for sample \#3     at ambient pressure ($p$ = 0) and at a pressure of 0.2 GPa. The data were taken in June 2021 (blue-shaded areas) and December 2023 (light green-shaded areas).}
    \label{fig:neffpressure}
\end{figure}
Further details of the magnetic response around the valence transition for sample \#3 taken from an outer part of the large sample are shown in Fig.~\ref{fig:neffpressure}. The figure includes data for $\mu_{\rm eff}$ as a function of temperature taken at ambient pressure ($p = 0$) and at a pressure of 0.2~GPa, yielding a reduction of the transition temperature of ($5 \pm 0.1$)~K, corresponding to a pressure dependence of ($25 \pm 1$)~K/GPa in agreement with literature results as discussed in Ref.~\cite{sarrao1998thermodynamics}.\\
To look for a thermal-history dependence, both the $p = 0$ and 0.2~GPa experiments were repeated after the sample had been cycled several times through the first-order valence transition. Whereas in the first experiments (June 2021) a narrow thermal hysteresis is revealed (blue-shaded areas), the hysteresis widens significantly (light green-shaded areas) in the later measurements (December 2023). Apparently, the magnetic hysteresis has widened significantly, and its midpoint has shifted to higher temperatures after the sample had been subject to multiple thermal cycles through the first-order transition. Based on Ref.~\cite{lawrence1996structure}, 
such a behavior can be expected when the amount of defects is increased. 
This, in fact, is expected as a consequence of the multiple thermal cycling through the first-order transition. 
\begin{figure}
    \centering
    \includegraphics[width=1\linewidth]{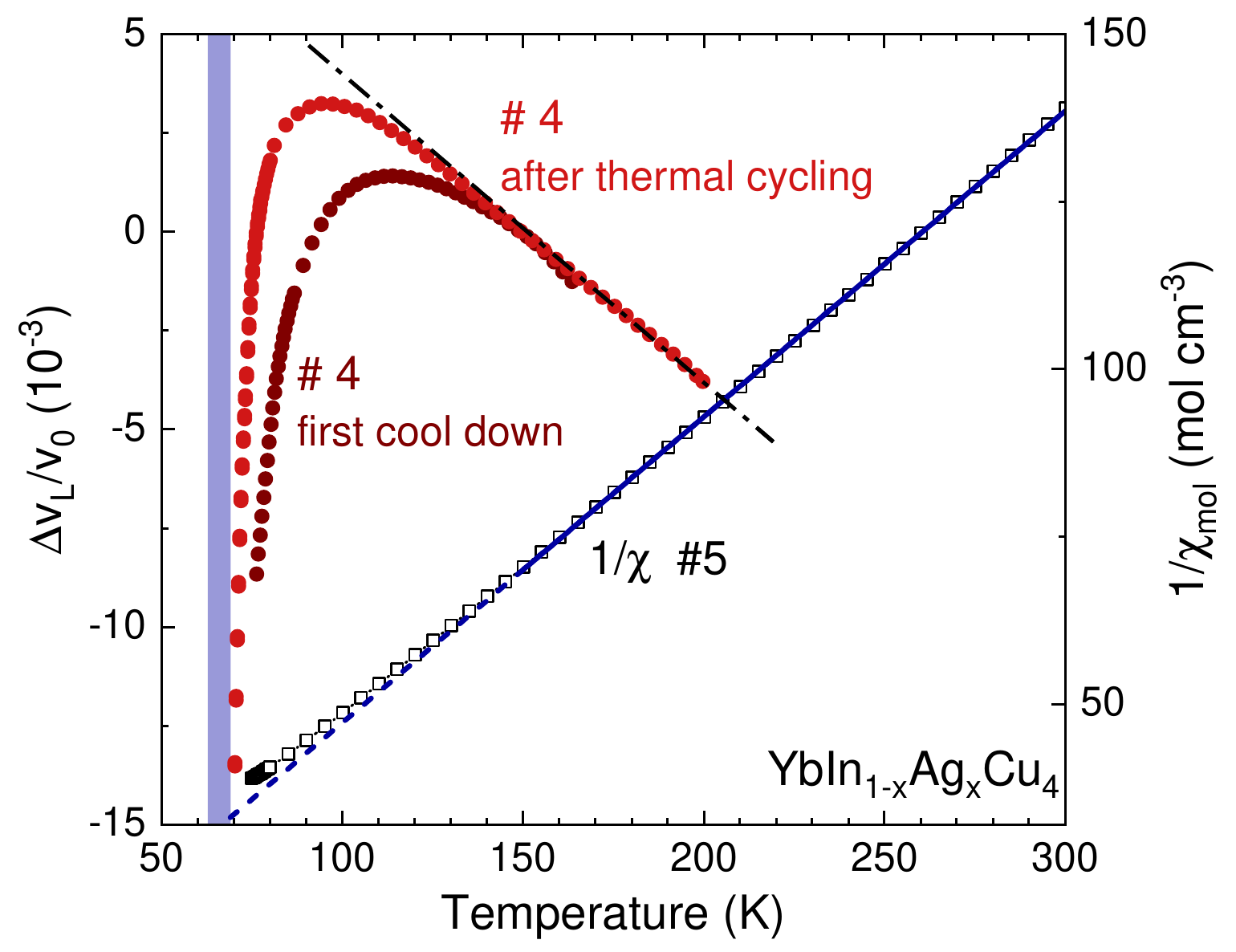}
    \caption{The inverse molar susceptibility (open black squares; right scale) for sample \#5 together with a Curie-Weiss fit to the data in the temperature range $150\,{\rm K}\leq T \leq 300\,\rm K$ (blue full line). Also shown is the temperature dependence of the longitudinal sound velocity $v_{\rm L}$($T$) for the neighboring sample (\#4, see inset of Fig.~\ref{fig:HC_Agseries}b) above the first-order valence transition (left scale). The dark red circles display data taken at the first cool-down, whereas the light red circles correspond to a subsequent cooling experiment. The blue shaded vertical bar indicates the temperature range, delimiting the array of transition temperatures for the neighboring sample \#5, inset of Fig.~\ref{fig:neff}. The broken black line represents the normal linear elastic background.}
    \label{fig:ultrasonic}
\end{figure}
Fig.~\ref{fig:ultrasonic} shows the inverse magnetic susceptibility of the outer piece \#5 (right scale) together with ultrasonic data measured on the neighboring single crystal \#4. The 1/$\chi_{\rm mol}$ data were fitted by using a Curie-Weiss expression for $150\,{\rm K}\leq T \leq 300\,\rm K$, corresponding to the blue full line. As indicated by the blue broken line, the experimental data start to deviate from the Curie-Weiss behavior below about 150~K, before they rapidly increase (not shown in Fig.~\ref{fig:ultrasonic}) below 70~K, indicating the onset of the valence transition. 
Fig.~\ref{fig:ultrasonic} also shows two data sets of the sound velocity $v_{\rm L}$ measured for sample  \#4 during the first cool down (dark red circles) and in a subsequent cool down (light red circles), i.e., after the crystal had been cooled through the valence transition. The longitudinal sound velocity $v_{\rm L}$ propagates along the [110] direction of the cubic crystal. The sound velocity $v_{\rm L}$ corresponds to the elastic mode $c_{\rm L}$ given by $c_{\rm L}$ = $v_{\rm L}^2 \cdot \rho_{\rm c}$, where $\rho_{\rm c}$ is the density of the material. The elastic mode $c_L$ is a linear combination of the three independent elastic modes for cubic symmetry, i.e., $c_{\rm L}$ = ($c_{11}$ + $c_{12}$ + 2 $c_{44}$)/2. Starting around 200 K, we first observe an almost linear increase in $c_{\rm L}$ upon cooling, which corresponds to the normal background elastic constant. In this temperature range, the susceptibility shows Curie-Weiss-like behavior. The $c_{\rm L}(T)$ data start to deviate from the straight line at around the same temperature, below which deviations from the Curie-Weiss-like behavior can be revealed in 1/$\chi_{\rm mol}$. Upon further cooling, c$_{\rm L}$ passes over a maximum around 110 K and becomes drastically reduced upon approaching the phase transition. This is accompanied by an increase in the ultrasonic attenuation (not shown). As a consequence, the ultrasonic signal could be followed only down to about 75~K, approximately 5 K above the onset of the transition. Upon further cooling down to 4.2~K, i.e., sufficiently below the transition temperature, the ultrasonic signal was found to recover (not shown here). A similar elastic behavior was found in Ref.~\cite{Zherlitsyn1999} on unsubstituted YbInCu$_4$ exhibiting a first-order transition. To look for effects related to the sample’s thermal history, the experiment was repeated (light red circles in Fig.~\ref{fig:ultrasonic}) after the sample had been cooled through the valence transition. Whereas there is no clear change in the normal elastic background, governing the data above 150~K, there are significant changes upon approaching the valence transition, manifesting themselves in a hardening of the elastic anomaly associated with the valence transition.

\subsubsection{Valence crossover samples close to the CEP} 
\begin{figure}[htbp]
    \centering
    \includegraphics[width=0.5\textwidth]{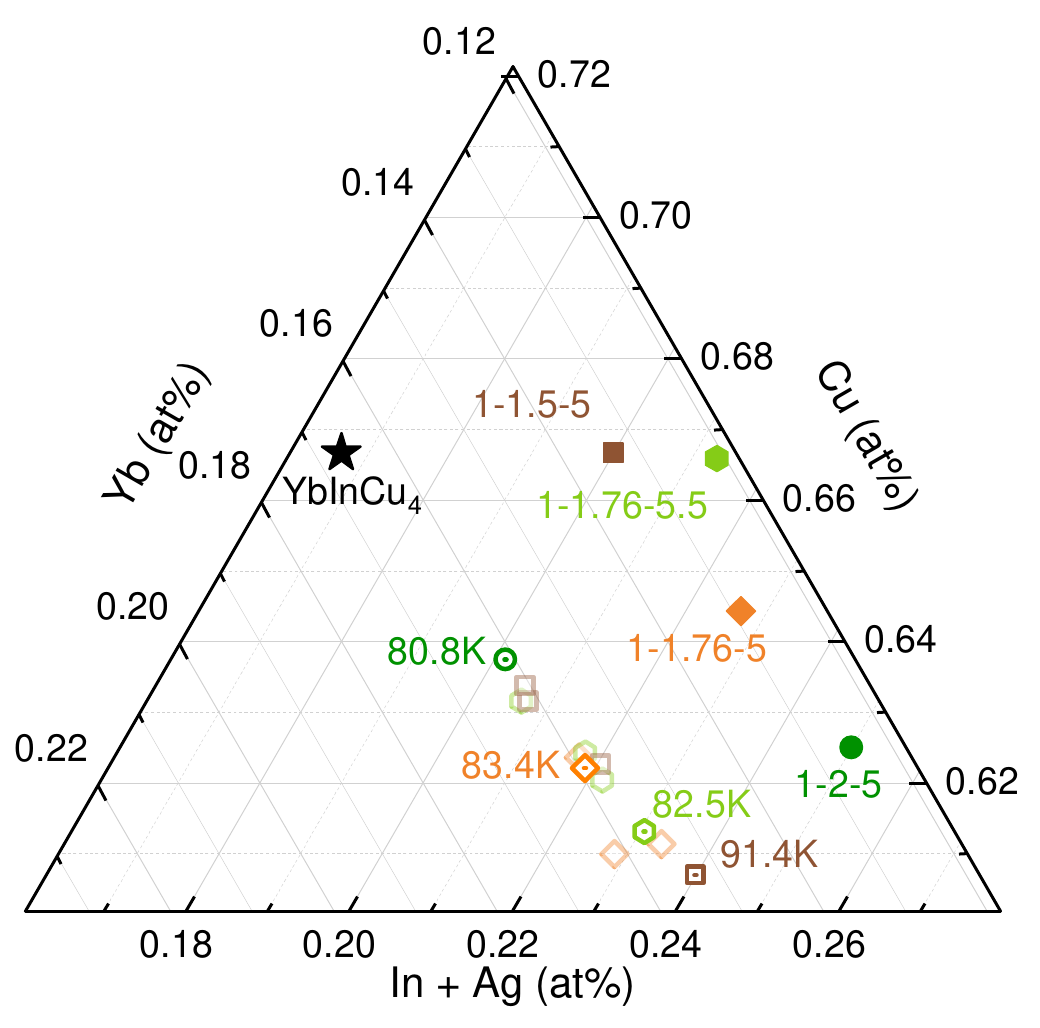}
    \caption{Enlarged view of the ternary phase diagram of Yb-(In+Ag)-Cu for $x_{\rm nom}$ = 0.20. The closed symbols indicate the initial composition of the melt while the corresponding open symbols are related to the chemical composition determined by EDX for the samples (open symbols with dots) where the heat capacity was measured (Fig.~\ref{fig:HC_ag}) and for other samples (pale symbols) from the same batches. Four different initial compositions of the melt are compared: 1-2-5 (green), 1-1.76-5 (orange), 1-1.76-5.5 (light green) and 1-1.5-5 (brown). }
    \label{fig:tenaerag}
\end{figure}

\begin{figure}[htbp]
    \centering
    \includegraphics[width=0.5\textwidth]{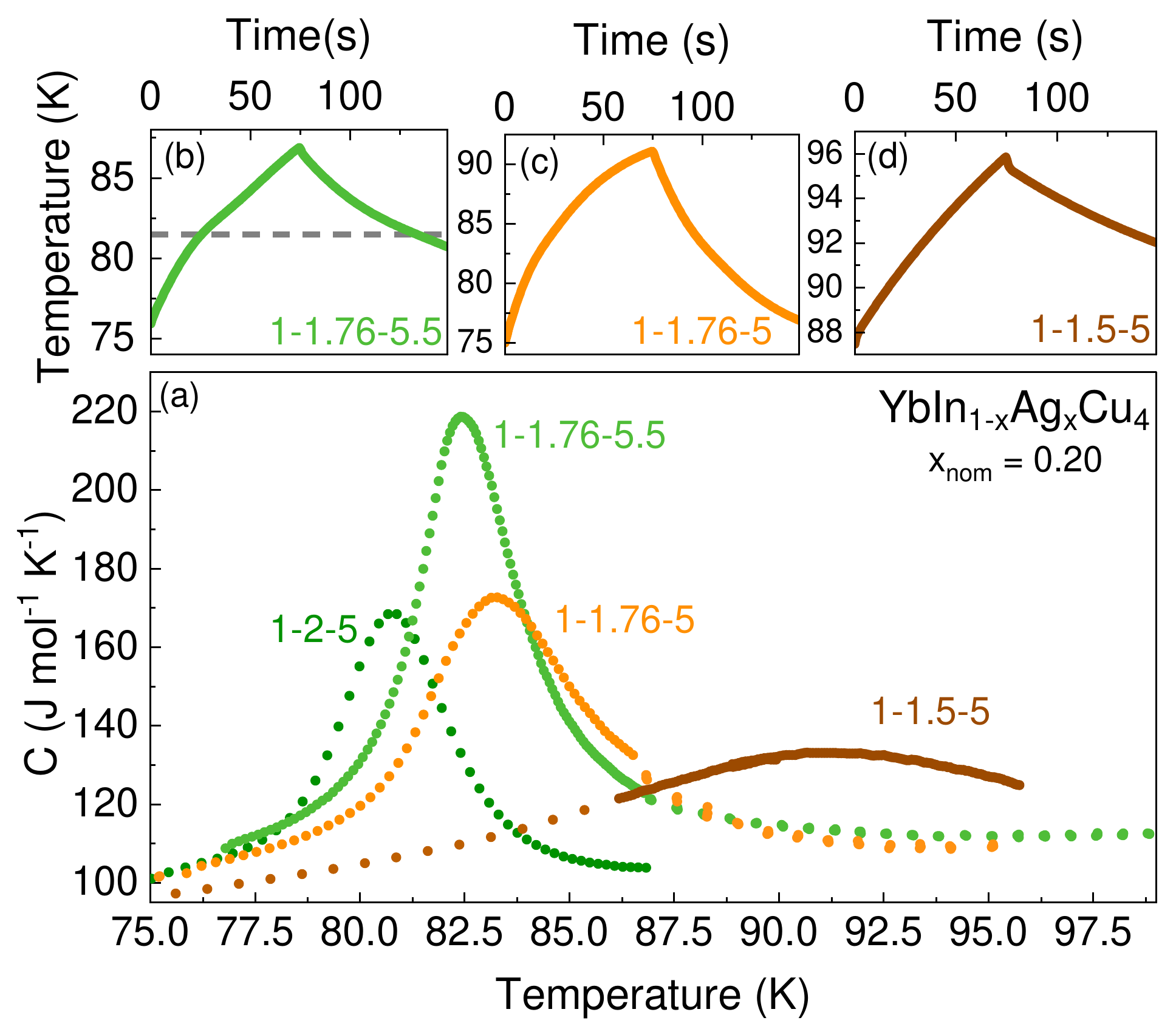}
    \caption{YbIn$_{1-x}$Ag$_x$Cu$_4$, $x_{\rm nom}$ = 0.20: (a) Heat capacity as function of temperature of samples grown from melts with different initial In-Cu ratio but constant Ag content of $x_{\rm nom}$ = 0.20 according to Yb-(In+Ag)-Cu. The heat pulses are shown which were measured on the 1-1.76-5.5 sample (b), on the 1-1.76-5 sample (c) and on the 1-1.5-5 sample (d). }
    \label{fig:HC_ag}
\end{figure}

\noindent
To investigate the CEP by applied hydrostatic He-gas pressure, samples are needed in the not too far distance on the low-pressure side of the phase transition, i.e., in the crossover regime. In Ag-substituted YbInCu$_4$, the CEP is assumed to be located at an Ag concentration of $x_{\rm nom} \approx 0.2$ \cite{sarrao1996evolution}. Therefore, we used this concentration as a starting point to investigate the influence of the variation of the In-Cu ratio at constant Ag concentration in the initial melt on the valence transition of the samples with respect to the order of their transition and their proximity to the putative CEP. 
In Fig.~\ref{fig:HC_Agseries}, we show heat capacity data for different Ag substitution levels grown from two different melt compositions 1-2-5 and 1-1.76-5. 
Now, we focus on samples with $x_{\rm nom} = 0.2$ and investigate in detail the influence of the In-Cu ratio on the valence transition using four samples with different initial melt compositions, Yb-(In+Ag)-Cu: 1-2-5, 1-1.76-5, 1-1.76-5.5 and 1-1.5-5, depicted as closed symbols in Fig.~\ref{fig:tenaerag}. Fig.~\ref{fig:HC_ag}(a) shows the heat capacity as function of temperature of the four samples with $x_{\rm nom}$ = 0.20  grown from these different initial compositions. The Ag concentration refers to the ratio Yb:In:Ag:Cu = 1:$y$(1-$x$):$yx$:5, where $y$ = 2, 1.76, 1.5 or Yb:In:Ag:Cu = 1:1.76(1-$x$):1.76$x$:5.5, respectively.  
Samples grown from the composition 1-2-5 as used in Ref.~\cite{sarrao1996ybin1} show the lowest transition temperature at 80.8\,K (green). The width of the peak (2.37 $\pm$ 0.03)\,K in the heat capacity is determined using a Gaussian fit.

\noindent
With decreasing In content in  the initial melt composition, the transition temperature shifts to higher temperatures and the transition gets broader. So the sample with less In but more Cu with the melt composition 1-1.76-5.5 shows a first-order transition at 82.5 K with a width of (2.30 $\pm$ 0.04)\,K (light green). A sample with less In 1-1.76-5 shows a transition at 83.4 K with a width of (3.5 $\pm$ 0.2)\,K  (orange). With further decreasing In content, the sample with 1-1.5-5.5 shows a broad crossover at the highest temperature 91.4 K and the largest width of (10.3 $\pm$ 0.5)\,K (brown). The character of the transition was determined using long heat pulses shown in Fig.~\ref{fig:HC_ag}(b-d). The two samples grown from 1-2-5 and 1-1.76-5.5 compositions (green and light green data) display a first-order transition while the data taken on the samples grown from 1-1.76-5 and 1-1.5-5 compositions (orange and brown) show a valence crossover.
It should be noted that this characterization is not always definite, for example the heat pulse measured on a $x = 0.20$ sample with 1-1.76-5 (orange) is not clearly continuous. This might be due to the fact that this sample is very close to the CEP and maybe small parts of it with a slightly different composition already have 1st-order characteristics. \\
The measured Ag content $x_{\rm EDX}$, summarized in Tab.~\ref{tab:Ag}, shows that the sample with the lowest transition temperature also has the lowest Ag content, whereas the variation is close to resolution limit of our EDX measurements.
To categorize the phase transitions, we examined samples grown from different melt stoichiometries, assuming that In is substituted by Ag according to Yb:In+Ag:Cu. In Fig.~\ref{fig:tenaerag}, besides the initial composition of the melt (closed symbols), the results of the EDX measurements on samples whose transition temperatures were compared by measuring their heat capacity (Fig.~\ref{fig:HC_ag}) are shown (open symbols). The pale symbols indicate the results of measurements that we made on samples from the same batches. 
It is noticeable that the Yb content remains constant for all compositions. However, the Ag+In-Cu ratio changes over the batches. From these data, it is apparent that not only the initial In-Cu ratio is the only criterion that determines the valence transition but also the Ag concentration which varies among one batch is important. This effect results in slightly different transition temperatures, meaning that $T_{\rm V}$ and $T^{\prime}_{\rm V}$  cannot be assigned to a fixed substitution level. 
\section{Discussion }
\begin{figure}[htbp]
    \centering
    \includegraphics[width=0.47\textwidth]{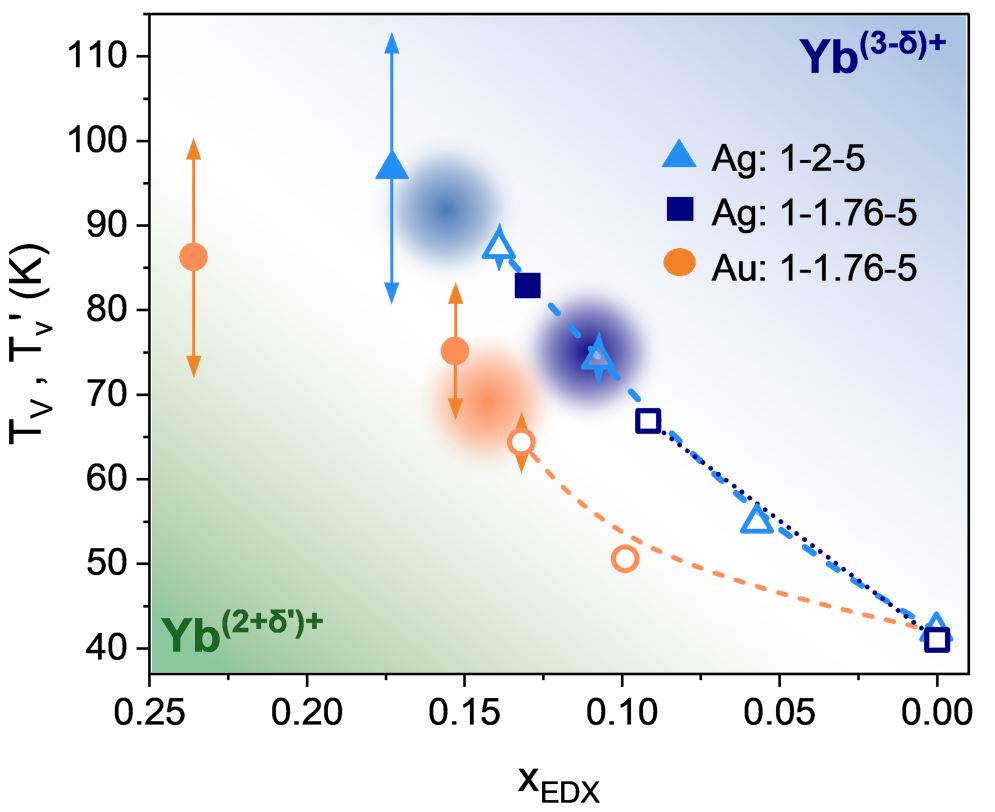}
    \caption{Temperature-substitution level phase diagram of YbInCu$_4$. 
    Transition temperature as function of the Ag- (triangles and cubes) or Au- (circles) concentration.
    Open symbols indicate a valence transition and closed symbols the valence crossover as determined from the presence of latent heat in the heat capacity. The dashed orange and blue lines are guides to the eyes and represents the line of the first-order valence transitions. The orange and blue circles mark the regions where the $2^{\rm nd}$ CEP is expected for the different substitution series. The arrows corresponds to half the value of the full width at half maximum of a Lorentz curves fitted to the HC data. In certain cases, the symbol exceeds the size of the displayed errors. }
    \label{fig:p-T-phasediagr}
\end{figure}

\noindent
We summarize our data of the chemically induced negative pressure on YbInCu$_4$ in a phase diagram as shown in Fig.~\ref{fig:p-T-phasediagr}.
The diagram presents data for two described Ag-substitution series with initial melt compositions of 1-2-5 and 1-1.76-5 as well as data for the Au-substitution series with an initial composition of 1-1.76-5.  With an increasing substitution level, the first-order valence transition (open symbols) shifts to higher temperatures until a valence crossover (closed symbols) is detected. The CEP is indicated here by the orange and blue areas, and appears to be dependent on the initial melt composition.
Usually, negative chemical pressure is achieved through substitution with larger atoms, so that the unit cell is enlarged, and the atoms (here Yb) are moved apart from each other. However, in the present case the lattice parameters do not change for substitution levels below $x = 0.30$. If we assume that the Ag/Au atoms are situated at the larger In site, negative chemical pressure is generated at the Yb site.
Through our investigations, we can therefore confirm the assumptions made in Ref.~\cite{sarrao1996evolution}.  
We measured an excess of In in all samples using EDX, which hints to In being incorporated into the Cu sites. With a constant lattice parameter, this creates positive pressure at the Yb site so that the low transition temperature around 40\,K is reached in the unsubstituted samples. This is in accordance with Ref.~\cite{loffert1999phase} where $T_{\rm V}\approx 40\,\rm K$ was found for samples with a slight Cu deficiency, while samples with a high Cu content show higher transition temperatures. Note that the SC analysis for the Ag-substituted samples did not indicate any In-Cu disorder. 
However, the substitution generates additional negative chemical pressure, so that the transition temperature increases with increasing substitution level. The substitution of In by smaller Ag and Au atoms while keeping the unit-cell volume constant ensures that Yb takes up more space in the unit cell, corresponding to a lower valence state. This only works up to a certain substituent concentration, so that the valence change no longer compensates the lattice change and the lattice parameter decreases for higher substitution levels. As outlined here, SC XRD reveals that Ag is probably not incorporated at the Cu sites, as this would lead to a larger $a$ lattice parameter.  

\section{Summary}\noindent
We have successfully grown YbInCu$_4$ single crystals from initial melts upon slightly varying their In-Cu ratio. Through these experiments, we obtained samples comparable to those reported in previous work \cite{sarrao1998thermodynamics} and found slight differences in the temperature $T_{\rm V}$, where the first-order valence transition occurs, ranging from $41\,{\rm K}\leq T_{\rm V}\leq 44\,\rm K$ depending on the stoichiometry of the initial melt. \\
Substitution experiments with Au on YbInCu$_4$ have been successfully performed. We found a high incorporation rate of Au in the crystals with $x_{\rm EDX} = 1.5\cdot x_{\rm nom}$. Samples with a substitution level of $x_{\rm nom}$ = 0.10 show a valence crossover at a temperature of $T^{\prime}_{\rm V}$ = 75\,K, which is lower than the valence-crossover temperatures of Ag-substituted samples. Through SC analysis, we were able to detect signatures of disorder in the samples, but were unable to identify which atoms were incorporated into which lattice sites.
\\
The substitution experiments with Ag show that $x_{\rm EDX} = 0.5\cdot x_{\rm nom}$, and that also here, the initial melt composition has an influence on the transition temperature. We found that melt compositions 1-2-5, 1-1.76-5.5 and 1-1.76-5 are well suited to grow samples close to a CEP. 
The critical substitution is reached for Ag-substitution levels between  $0.2\leq x_{\rm nom}\leq 0.27$ with $83 \,{\rm K}\leq T_{\rm V}\leq 98\,\rm K$ depending on the initial In-Cu ratio of the melt. 
\\
Multiple thermal cycles through the first-order transition, as demonstrated for the well characterized sample \#3, lead to a significant modification of the magnetic properties. These experimental findings clearly demonstrate the strong influence of defects on the valence transition in substituted YbInCu$_4$.
\\
Our crystal growth experiments show that in case of Au-substitution as well as in case of Ag-substitution we are able to fine tune the initial melt stoichiometry with respect to the level of substitution and the initial In-Cu ratio of the melt to obtain samples with defined properties close to the CEP. These samples are now subject to in-depth investigations to study possible critical elasticity in these intermetallic materials.

\begin{acknowledgments}\noindent
We thank C. Geibel for valuable discussions and T. F\"orster for technical support. We acknowledge funding by the Deutsche Forschungsgemeinschaft (DFG, German Research Foundation) via the TRR 288 (422213477, projects A01 and A03).
\end{acknowledgments}

\cleardoublepage
\newpage
\noindent
\textbf{\Large Supplemental Material}
\renewcommand{\thefigure}{S\arabic{figure}}
\setcounter{figure}{0}

\subsection{YbInCu$_4$ \label{sec:YbInCu4-HC}}\noindent
\begin{figure}[htpb]
    \centering
\includegraphics[width=0.47\textwidth]{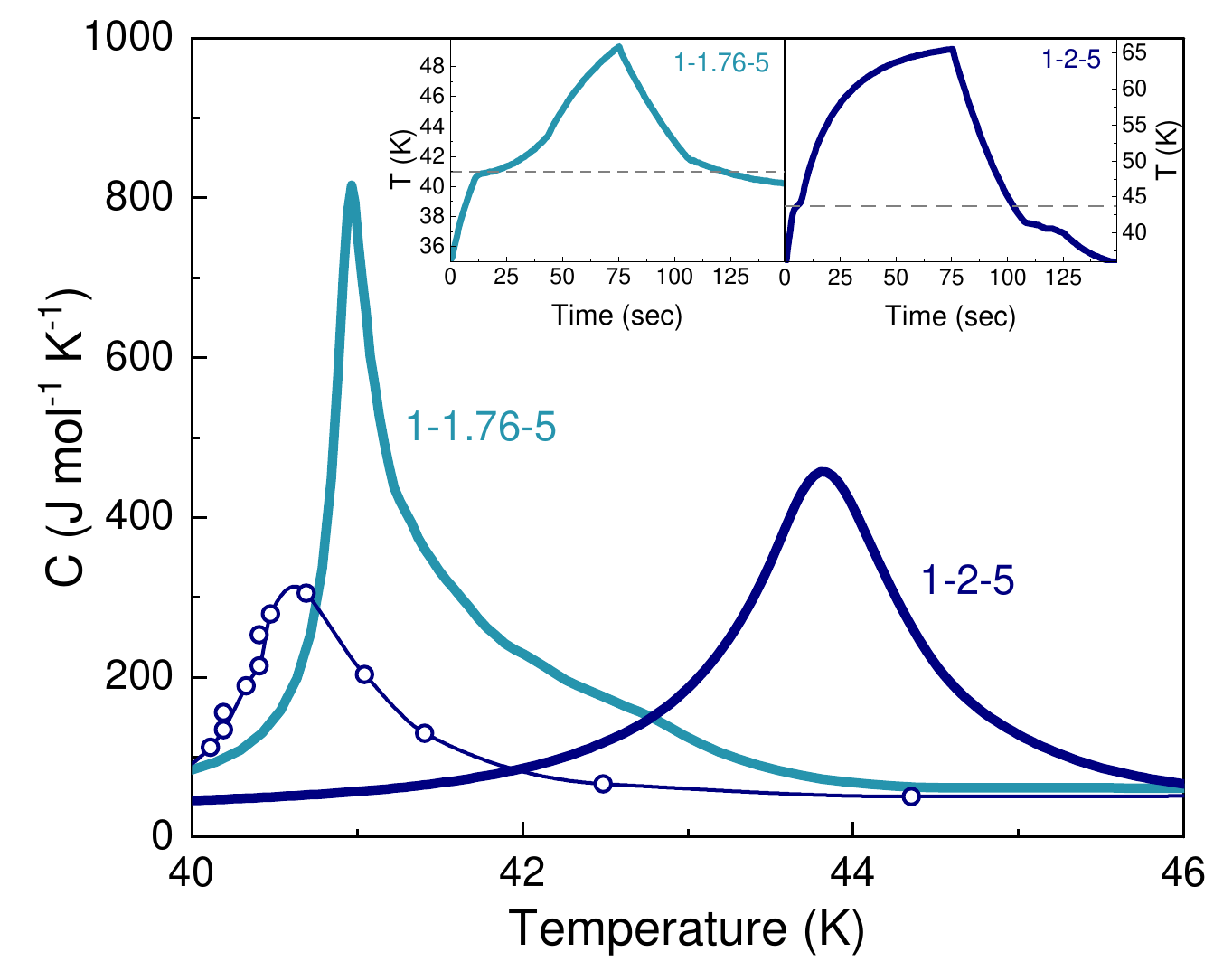}
    \caption{YbInCu$_4$ heat capacity calculated from the heating curve of the heat pulses shown in the inset. The sample grown from the initial composition 1-1.76-5 (teal) shows a lower transition temperature of $T_{\rm V}$ = 41\,K with a broad high-temperature shoulder. The sample grown from initial composition 1-2-5 (dark blue, closed symbols) displayes the  transition at higher temperatures with a peak at $T_{\rm V}$ = 44\,K. 
    For comparison, data reported by \cite{sarrao1998thermodynamics} (open symbols, samples grown from 1-2-5 composition) are shown which display the valence transition with a peak at $T_{\rm V}\approx$ 40.5\,K. }
    \label{fig:HC_unsub}
\end{figure}\noindent
To find the starting composition that produces samples of pure YbInCu$_4$  with low disorder, we conducted various crystal growth experiments and characterized the samples using heat capacity measurements.
A sharp peak in the heat capacity is associated with a low disorder of the sample. In addition, the heat capacity can be used to deduce the order of the phase transition. A first-order phase transition is associated with the occurrence of latent heat at the transition, which is visible as a kink in the heating pulse signal. 
Fig.~\ref{fig:HC_unsub} shows the temperature dependence of the heat capacity of two samples from different growths, where the peak positions and shapes are compared. Both samples were grown from In-Cu flux, with different initial compositions Yb:In:Cu = 1:2:5 (1-2-5, dark blue symbols, composition formerly used in Ref.~\cite{sarrao1996ybin1}) and Yb:In:Cu = 1:1.76:5 (1-1.76-5, teal symbols). 
There are small differences in the transition temperature: 
The sample from the In-rich initial composition, 1-2-5, shows a clear peak at $T_{\rm V} = $ 44\,K, while the sample grown from the initial melt with less In, 1-1.76-5, shows a sharp peak at $T_{\rm V} = 41$\,K. 
Besides this sharp peak, the data recorded on the 1-1.76-5 
sample show a broad high-temperature shoulder extending over a range of $\approx$ 2 K which is probably caused by the presence of regions with tiny variations of the In-Cu ratio. The plotted data was calculated from the heat pulses shown in the inset of Fig.~\ref{fig:HC_unsub}. The occurrence of latent heat (kink in continuous curve) characterizes the transition as first-order. 
It should be noted that due to the hysteresis of a first-order phase transition, the transition temperature is not perfectly defined. For this reason, we have always used the heating curve for this analysis. 
To conclude, in our crystal growth experiments we could achieve samples with comparable properties like those reported in previous work \cite{sarrao1998thermodynamics}. 
The thorough screening of different samples by heat capacity measurements yielded, that often samples show more than one transition caused by a slightly varying In-Cu ratio within one crystal. Therefore, identifying the samples with the lowest amount of disorder by analyzing the width of the peak in the heat capacity is hindered in case of larger samples.

\subsection{Results of the SC-XRD on Au and Ag substituted YbInCu$_4$ \label{sec:SCXRD-Au}}\noindent
The crystallographic data and details of the SC-XRD for Au-substituted samples are summarized in Tab.~\ref{tab:sc-au} and for Ag-substituted samples in Tab.~\ref{tab:sc-ag}.
While the refinement fails in the Ag substituted samples due to too small difference of the atomic scattering factors of the elements In and Ag, two models can be obtained for the disorder in the Au-substituted samples.  In the model A we assumed (Yb(In$_{1-x}$Au$_x$)(Cu$_{1-y}$Au$_y$)$_4$), while in model B we assumed (Yb(In$_{1-x}$Au$_x$)(Cu$_{1-y}$In$_y$)$_4$).

\begin{table*}[]
    \centering
    \begin{tabular}{|c|cccc|}
    \hline
         sample&  \multicolumn{2}{c|}{\textbf{x$_{\rm nom}$ = 0.1}}&  \multicolumn{2}{c|}{\textbf{x$_{\rm nom}$ = 0.3}} \\
         &Model A & \multicolumn{1}{c|}{Model B} & Model A& Model B \\ 
        \hline
        Chemical formula &Au$_{0.23}$Cu$_{3.86}$In$_{0.91}$Yb &\multicolumn{1}{c|}{Au$_{0.08}$Cu$_{3.58}$In$_{1.35}$Yb} & Au$_{0.52}$Cu$_{3.84}$In$_{0.65}$Yb&Au$_{0.35}$Cu$_{3.49}$In$_{1.16}$Yb \\ 
        M$_r$ &567.76 &\multicolumn{1}{c|}{570.67} &593.48 & 596.80\\
        Crystal system  &\multicolumn{2}{c|}{Cubic }&\multicolumn{2}{c|}{Cubic} \\
        Space group &\multicolumn{2}{c|}{$F\overline{4}3m$ }& \multicolumn{2}{c|}{$F\overline{4}3m$}\\
        Temperature (K) &\multicolumn{2}{c|}{240 }&\multicolumn{2}{c|}{240 }\\
        a (\AA) &\multicolumn{2}{c|}{7.17256(7) }&\multicolumn{2}{c|}{7.1499(3) }  \\
        V (\AA$^3$) &\multicolumn{2}{c|}{369.00(1) } & \multicolumn{2}{c|}{365.51(5) }\\
        Z & \multicolumn{2}{c|}{4}&\multicolumn{2}{c|}{4} \\
        F(000) & 978&\multicolumn{1}{c|}{984}&1016&1023 \\
        D$_x$ (Mg m$^{-3}$) &10.220&\multicolumn{1}{c|}{10.272}&10.785& 1023\\
        Radiation type & \multicolumn{2}{c|}{Mo K$\alpha$}&\multicolumn{2}{c|}{Mo K$\alpha$ }\\
        $\mu$ (mm$^{-1}$) &61.29&\multicolumn{1}{c|}{56.50}&71.78& 66.19\\
        Crystal size (mm) &\multicolumn{2}{c|}{0.01 × 0.01 × 0.003}&\multicolumn{2}{c|}{0.01 × 0.01 × 0.01}\\
        Crystal shape, color & \multicolumn{2}{c|}{Plate, metallic luster silver}&\multicolumn{2}{c|}{Plate, metallic luster silver} \\
        No. of measured,& \multicolumn{2}{c|}{2612, 101, 101}&\multicolumn{2}{c|}{3611, 96, 96} \\
        independent and  & \multicolumn{2}{c|}{}&\multicolumn{2}{c|}{} \\
        observed $[$I $>$ 2$\sigma$(I)$]$  
         & \multicolumn{2}{c|}{}&\multicolumn{2}{c|}{} \\
        reflections & \multicolumn{2}{c|}{}&\multicolumn{2}{c|}{} \\
        R$_{\rm int}$ &\multicolumn{2}{c|}{0.093}&\multicolumn{2}{c|}{0.105} \\
        2$\Theta_{\rm max}(\circ)$ &\multicolumn{2}{c|}{68.2}&\multicolumn{2}{c|}{68.4}\\
       R(F)$^a$, wR(F$^2$)$^b$, GooF$^c$ &0.018, 0.055, 1.23 &\multicolumn{1}{c|}{0.018, 0.053, 1.19}&0.016, 0.046, 1.46&0.016, 0.047, 1.43\\
        No. of parameters &9&\multicolumn{1}{c|}{9}&9&9 \\
        Extinction coefficient &0.0024(7)&\multicolumn{1}{c|}{0.0025(7)}&0.0027(5)&0.0029(6)\\
        $\Delta \rho_{\rm max}$, $\Delta \rho_{\rm min}$ (e \AA $^{-3}$) &0.94, -1.91&\multicolumn{1}{c|}{0.95, -1.85}&1.18, -0.97&1.13, -0.91\\
        Absolute structure parameter &-0.11(5) &\multicolumn{1}{c|}{-0.15(5)}&-0.16(15)&0.02(15) \\
        \hline
    \end{tabular}
    \caption{The crystallographic data and details of the diffraction experiments for Au-substituted samples.}
    \label{tab:sc-au}
\end{table*}

\begin{table}[]
    \centering
    \begin{tabular}{|c|c|c|}
    \hline
        sample  & \textbf{x$_{\rm nom}$ = 0.0}& \textbf{x$_{\rm nom}$ = 0.20}\\
        \hline
        \hline
        Chemical formula &Cu$_4$InYb & Cu$_4$InYb\\
        \hline
        M$_r$ &542.02 & 542.02\\
        \hline
        Crystal system  &Cubic &Cubic \\
        \hline
        Space group &$F\overline{4}3m$ & $F\overline{4}3m$\\
        \hline
        Temperature (K) &240 &240 \\
        \hline
        a (\AA) & 7.15631(8)&7.16095(9)  \\
        \hline
        V (\AA$^3$) &366.49(1) & 367.21(1)\\
        \hline
        Z & 4&4 \\
        \hline
        F(000) &940 &940 \\
        \hline
        D$_x$ (Mg m$^{-3}$) &9.823 &9.823 \\
        \hline
        Radiation type & Mo K$\alpha$&Mo K$\alpha$ \\
        \hline
        $\mu$ (mm$^{-1}$) &53.97 &53.97 \\
        \hline
        Crystal size (mm) &0.03 × 0.03  & 0.03 × 0.02 \\
         & × 0.01 &  × 0.01\\
         \hline
        Crystal shape & Plate, m&Plate \\
        \hline
        Crystal color &  metallic luster & metallic luster 
        \\
                      & silver&silver\\
        \hline
        No. of measured, &2072, 96, 96 &2202, 95, 95 \\
        independent and   & & \\
        observed $[$I $>$ 2$\sigma$(I)$]$  & & \\
        reflections & & \\
        \hline
        R$_{\rm int}$ &0.056 &0.064 \\
        \hline
        2$\Theta_{\rm max}(\circ)$ &68.4 & 66.8\\
        \hline
       R(F)$^a$, wR(F$^2$)$^b$,  &0.014, 0.038,  & 0.014, 0.037\\
       \hline
        GooF$^c$ & 1.24 &  1.37\\
        \hline
        No. of parameters &7 &7 \\
        \hline
        Extinction coefficient &0.0011(3) & 0.0015(4)\\
        \hline
        $\Delta \rho_{\rm max}$, $\Delta \rho_{\rm min}$ (e \AA $^{-3}$) &0.91, -1.43 & 1.21, -1.03\\
        \hline
        Absolute structure &0.00(2) &-0.01(3) \\
        parameter & & \\
        \hline
    \end{tabular}
    \caption{The crystallographic data and details of the diffraction experiments for Ag-substituted samples.}
    \label{tab:sc-ag}
\end{table}

\subsection{Laue analysis \label{sec:Laue}} 
\begin{figure}[htpb]
    \centering
\includegraphics[width=0.47\textwidth]{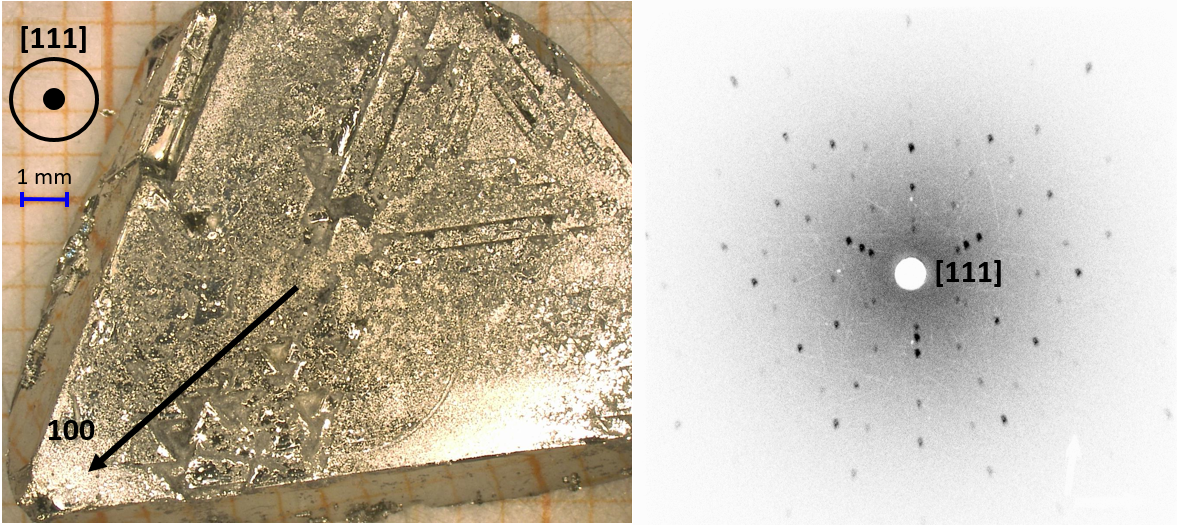}
    \caption{YbIn$_{1-x}$Ag$_x$Cu$_4$: Oriented single crystal on mm grid  and Laue pattern. }
    \label{fig:Laue}
\end{figure}

\noindent
An oriented single crystal and the Laue analysis is shown in Fig.~\ref{fig:Laue}. 
The figure shows the sample whose pieces are discussed in section \ref{chap:HC_Ag} and \ref{chap:valencetransition}. The sample was aligned and cut so that measurements could be taken along certain crystallographic directions.  

\subsection{Magnetic susceptibility of Au-substituted samples \label{sec:Ausus}}
\begin{figure}[htbp]
    \centering
    \includegraphics[width=0.5\textwidth]{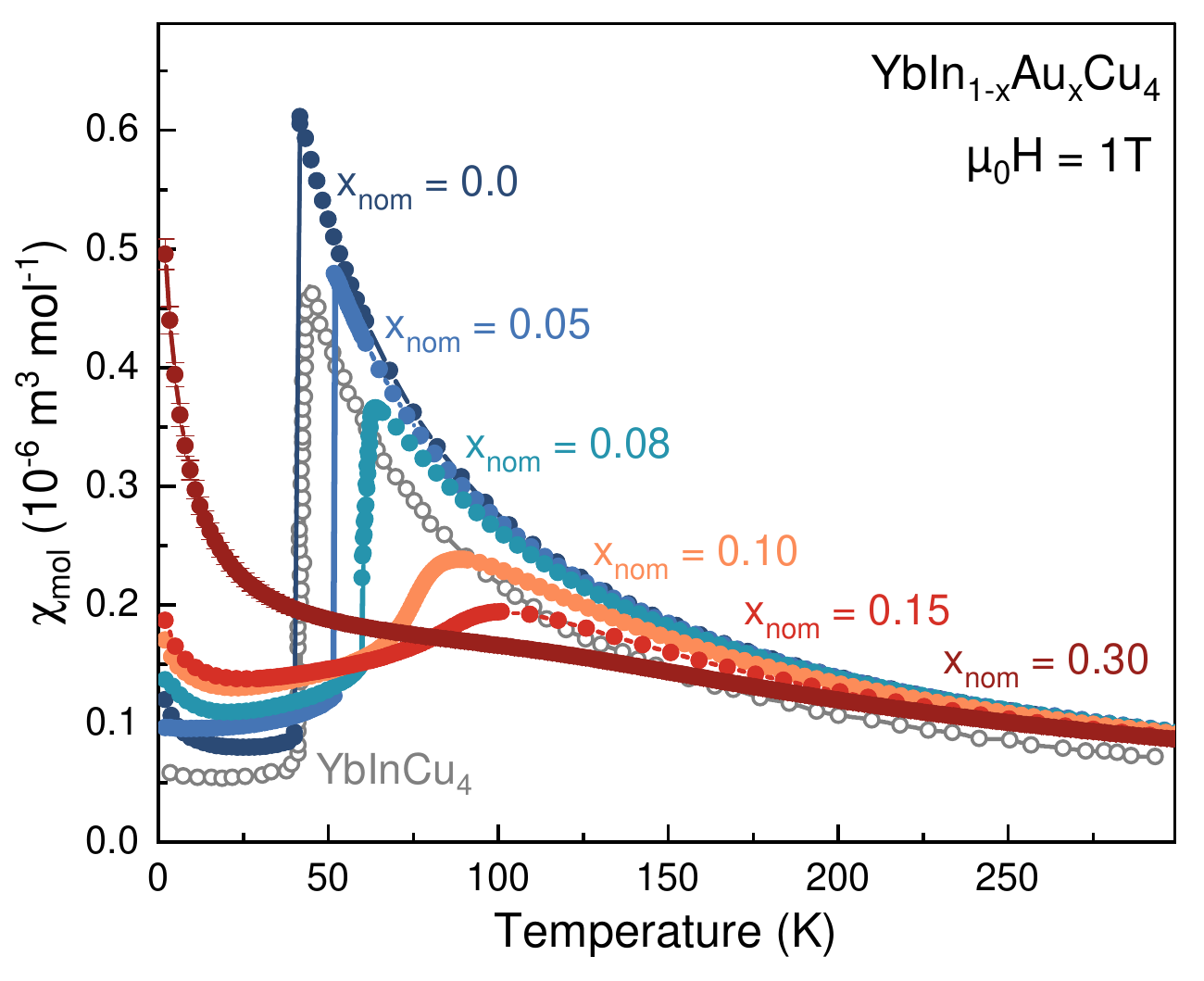}
    \caption{YbIn$_{1-x}$Au$_x$Cu$_4$: Magnetic susceptibility as a function of temperature for several Au concentrations. Heating and cooling curves are plotted in each case. The data of YbInCu$_4$ (gray) is taken from \cite{sarrao1996ybin1}.}
    \label{fig:sus}
\end{figure}

\noindent
The magnetic susceptibility as a function of temperature was measured in a field of $\mu_0H = 1\,\rm T$ and is shown in Fig.~\ref{fig:sus}. At high temperatures, the magnetic susceptibility follows the Curie-Weiss law. Upon cooling below the transition temperature, the susceptibility drops to a value of nearly compensated magnetic moments, which is associated with the divalent Yb state. Consistent with the heat capacity, the valence transition shifts to higher temperatures with increasing Au-substitution level. With higher Au content, the transition broadens and the absolute value of the susceptibility becomes smaller. The sample with $x_{\rm nom}$ = 0.30 shows Curie-Weiss behavior down to $T = 230\,\rm K$.\\

\subsection{Magnetic susceptibility of Ag-substituted samples \label{sec:Laue}}
\begin{figure}[htpb]
    \centering
\includegraphics[width=0.47\textwidth]{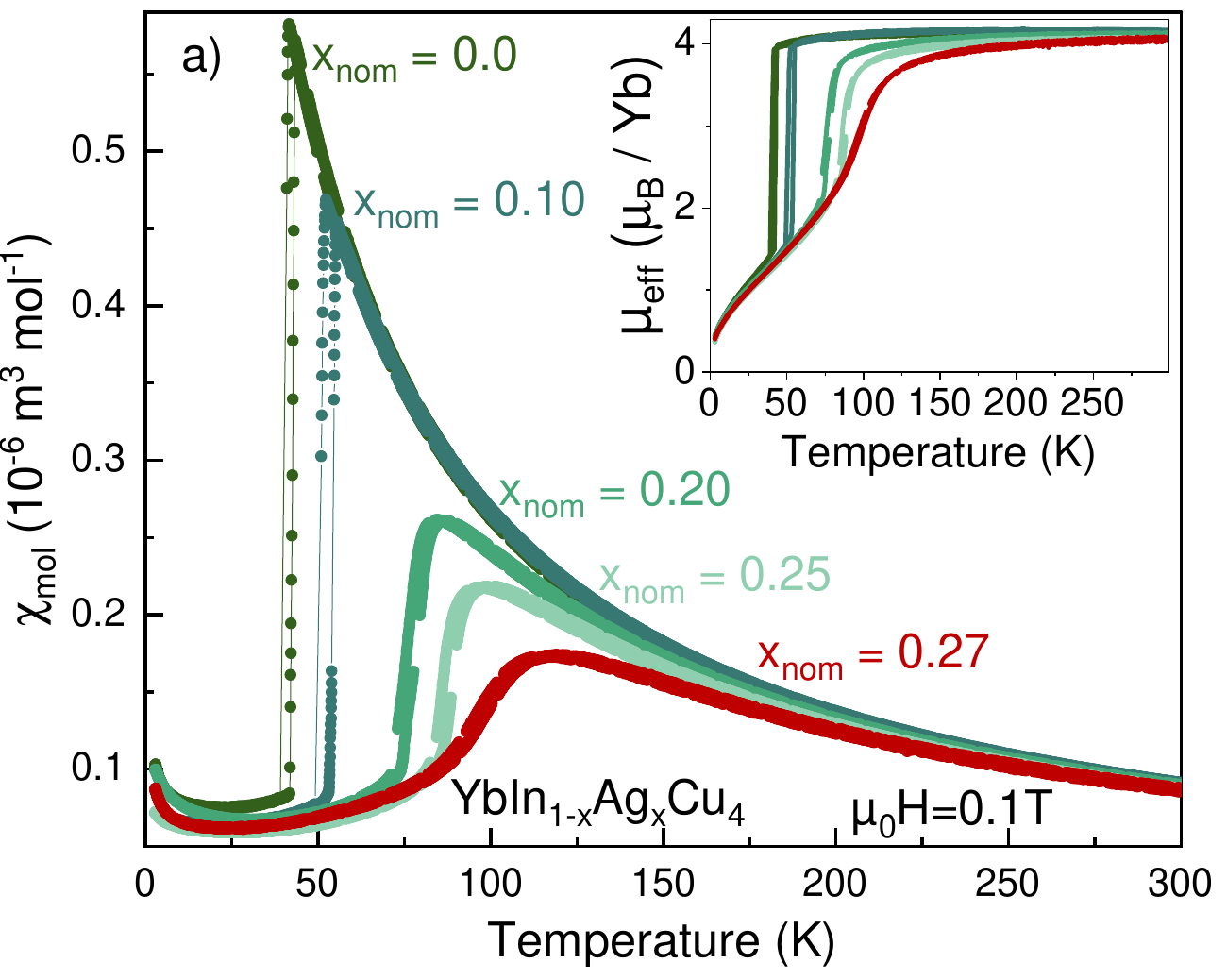}
\includegraphics[width=0.47\textwidth]{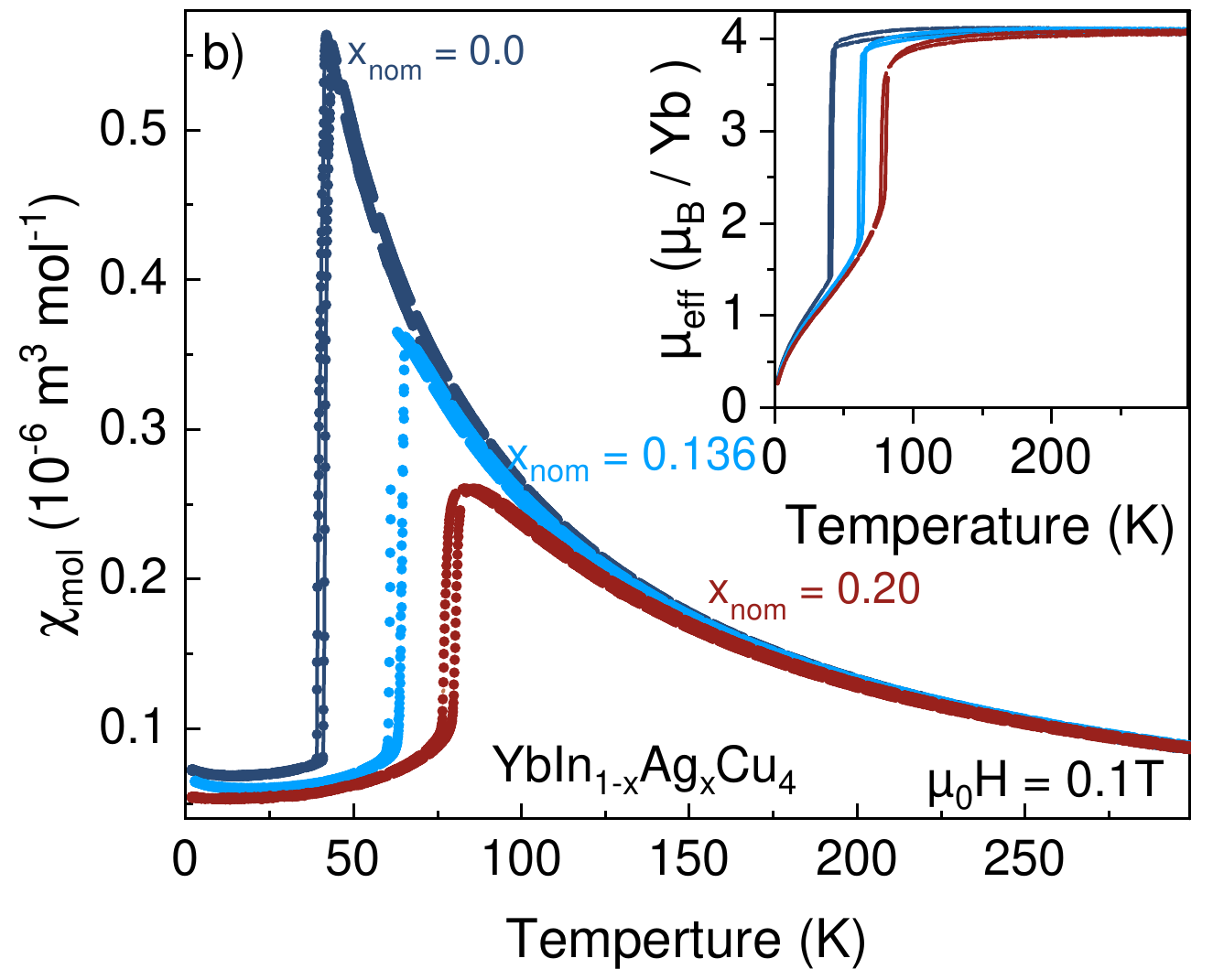}
    \caption{Magnetic susceptibility and effective magnetic moment $\mu_{\rm eff}$ (inset) measured on YbIn$_{1-x}$Ag$_x$Cu$_4$ samples grown from a melt with the initial composition (a) 1-2-5 and (b) 1-1.76-5. }
    \label{fig:ChiTAg}
\end{figure}

\noindent
The magnetic susceptibility as function of the temperature for different initial compositions of the melt and different substitution levels is shown in Fig. \ref{fig:ChiTAg}. The applied magnetic field is $\mu_0H$ = 0.1T. The inset shows the effective moment $\mu_{\rm eff}$ calculated from the data.
\cleardoublepage
\bibliography{Lit}

\begin{thebibliography}{35}%
\makeatletter
\providecommand \@ifxundefined [1]{%
 \@ifx{#1\undefined}
}%
\providecommand \@ifnum [1]{%
 \ifnum #1\expandafter \@firstoftwo
 \else \expandafter \@secondoftwo
 \fi
}%
\providecommand \@ifx [1]{%
 \ifx #1\expandafter \@firstoftwo
 \else \expandafter \@secondoftwo
 \fi
}%
\providecommand \natexlab [1]{#1}%
\providecommand \enquote  [1]{``#1''}%
\providecommand \bibnamefont  [1]{#1}%
\providecommand \bibfnamefont [1]{#1}%
\providecommand \citenamefont [1]{#1}%
\providecommand \href@noop [0]{\@secondoftwo}%
\providecommand \href [0]{\begingroup \@sanitize@url \@href}%
\providecommand \@href[1]{\@@startlink{#1}\@@href}%
\providecommand \@@href[1]{\endgroup#1\@@endlink}%
\providecommand \@sanitize@url [0]{\catcode `\\12\catcode `\$12\catcode
  `\&12\catcode `\#12\catcode `\^12\catcode `\_12\catcode `\%12\relax}%
\providecommand \@@startlink[1]{}%
\providecommand \@@endlink[0]{}%
\providecommand \url  [0]{\begingroup\@sanitize@url \@url }%
\providecommand \@url [1]{\endgroup\@href {#1}{\urlprefix }}%
\providecommand \urlprefix  [0]{URL }%
\providecommand \Eprint [0]{\href }%
\providecommand \doibase [0]{https://doi.org/}%
\providecommand \selectlanguage [0]{\@gobble}%
\providecommand \bibinfo  [0]{\@secondoftwo}%
\providecommand \bibfield  [0]{\@secondoftwo}%
\providecommand \translation [1]{[#1]}%
\providecommand \BibitemOpen [0]{}%
\providecommand \bibitemStop [0]{}%
\providecommand \bibitemNoStop [0]{.\EOS\space}%
\providecommand \EOS [0]{\spacefactor3000\relax}%
\providecommand \BibitemShut  [1]{\csname bibitem#1\endcsname}%
\let\auto@bib@innerbib\@empty
\bibitem [{\citenamefont {B{\"o}hmer}\ and\ \citenamefont
  {Meingast}(2016)}]{bohmer2016electronic}%
  \BibitemOpen
  \bibfield  {author} {\bibinfo {author} {\bibfnamefont {A.~E.}\ \bibnamefont
  {B{\"o}hmer}}\ and\ \bibinfo {author} {\bibfnamefont {C.}~\bibnamefont
  {Meingast}},\ }\bibfield  {title} {\bibinfo {title} {Electronic nematic
  susceptibility of iron-based superconductors},\ }\href
  {https://doi.org/https://doi.org/10.1016/j.crhy.2015.07.001} {\bibfield
  {journal} {\bibinfo  {journal} {Comptes Rendus Physique}\ }\textbf {\bibinfo
  {volume} {17}},\ \bibinfo {pages} {90} (\bibinfo {year} {2016})}\BibitemShut
  {NoStop}%
\bibitem [{\citenamefont {Yao}\ \emph {et~al.}(2022)\citenamefont {Yao},
  \citenamefont {Willa}, \citenamefont {Lacmann}, \citenamefont {Souliou},
  \citenamefont {Frachet}, \citenamefont {Willa}, \citenamefont {Merz},
  \citenamefont {Weber}, \citenamefont {Meingast}, \citenamefont {Heid} \emph
  {et~al.}}]{yao2022electronic}%
  \BibitemOpen
  \bibfield  {author} {\bibinfo {author} {\bibfnamefont {Y.}~\bibnamefont
  {Yao}}, \bibinfo {author} {\bibfnamefont {R.}~\bibnamefont {Willa}}, \bibinfo
  {author} {\bibfnamefont {T.}~\bibnamefont {Lacmann}}, \bibinfo {author}
  {\bibfnamefont {S.-M.}\ \bibnamefont {Souliou}}, \bibinfo {author}
  {\bibfnamefont {M.}~\bibnamefont {Frachet}}, \bibinfo {author} {\bibfnamefont
  {K.}~\bibnamefont {Willa}}, \bibinfo {author} {\bibfnamefont
  {M.}~\bibnamefont {Merz}}, \bibinfo {author} {\bibfnamefont {F.}~\bibnamefont
  {Weber}}, \bibinfo {author} {\bibfnamefont {C.}~\bibnamefont {Meingast}},
  \bibinfo {author} {\bibfnamefont {R.}~\bibnamefont {Heid}}, \emph {et~al.},\
  }\bibfield  {title} {\bibinfo {title} {An electronic nematic liquid in
  {B}a{N}i$_2${A}s$_2$},\ }\href
  {https://www.nature.com/articles/s41467-022-32112-7} {\bibfield  {journal}
  {\bibinfo  {journal} {Nature communications}\ }\textbf {\bibinfo {volume}
  {13}},\ \bibinfo {pages} {4535} (\bibinfo {year} {2022})}\BibitemShut
  {NoStop}%
\bibitem [{\citenamefont {Li}\ \emph {et~al.}(2022)\citenamefont {Li},
  \citenamefont {Garst}, \citenamefont {Schmalian}, \citenamefont {Ghosh},
  \citenamefont {Kikugawa}, \citenamefont {Sokolov}, \citenamefont {Hicks},
  \citenamefont {Jerzembeck}, \citenamefont {Ikeda}, \citenamefont {Hu} \emph
  {et~al.}}]{li2022elastocaloric}%
  \BibitemOpen
  \bibfield  {author} {\bibinfo {author} {\bibfnamefont {Y.-S.}\ \bibnamefont
  {Li}}, \bibinfo {author} {\bibfnamefont {M.}~\bibnamefont {Garst}}, \bibinfo
  {author} {\bibfnamefont {J.}~\bibnamefont {Schmalian}}, \bibinfo {author}
  {\bibfnamefont {S.}~\bibnamefont {Ghosh}}, \bibinfo {author} {\bibfnamefont
  {N.}~\bibnamefont {Kikugawa}}, \bibinfo {author} {\bibfnamefont {D.~A.}\
  \bibnamefont {Sokolov}}, \bibinfo {author} {\bibfnamefont {C.~W.}\
  \bibnamefont {Hicks}}, \bibinfo {author} {\bibfnamefont {F.}~\bibnamefont
  {Jerzembeck}}, \bibinfo {author} {\bibfnamefont {M.~S.}\ \bibnamefont
  {Ikeda}}, \bibinfo {author} {\bibfnamefont {Z.}~\bibnamefont {Hu}}, \emph
  {et~al.},\ }\bibfield  {title} {\bibinfo {title} {Elastocaloric determination
  of the phase diagram of {S}r$_2${R}u{O}$_4$},\ }\href
  {https://doi.org/10.1038/s41586-022-04820-z} {\bibfield  {journal} {\bibinfo
  {journal} {Nature}\ }\textbf {\bibinfo {volume} {607}},\ \bibinfo {pages}
  {276} (\bibinfo {year} {2022})}\BibitemShut {NoStop}%
\bibitem [{\citenamefont {Xiao}\ \emph {et~al.}(2021)\citenamefont {Xiao},
  \citenamefont {Borisov}, \citenamefont {Gorgen-Lesseux}, \citenamefont
  {Rommel}, \citenamefont {Song}, \citenamefont {Maita}, \citenamefont
  {Aindow}, \citenamefont {Valentí}, \citenamefont {Canfield},\ and\
  \citenamefont {Lee}}]{Xiao2021}%
  \BibitemOpen
  \bibfield  {author} {\bibinfo {author} {\bibfnamefont {S.}~\bibnamefont
  {Xiao}}, \bibinfo {author} {\bibfnamefont {V.}~\bibnamefont {Borisov}},
  \bibinfo {author} {\bibfnamefont {G.}~\bibnamefont {Gorgen-Lesseux}},
  \bibinfo {author} {\bibfnamefont {S.}~\bibnamefont {Rommel}}, \bibinfo
  {author} {\bibfnamefont {G.}~\bibnamefont {Song}}, \bibinfo {author}
  {\bibfnamefont {J.~M.}\ \bibnamefont {Maita}}, \bibinfo {author}
  {\bibfnamefont {M.}~\bibnamefont {Aindow}}, \bibinfo {author} {\bibfnamefont
  {R.}~\bibnamefont {Valentí}}, \bibinfo {author} {\bibfnamefont {P.~C.}\
  \bibnamefont {Canfield}},\ and\ \bibinfo {author} {\bibfnamefont {S.-W.}\
  \bibnamefont {Lee}},\ }\bibfield  {title} {\bibinfo {title} {Pseudoelasticity
  of {S}r{N}i$_2${P}$_2$ micropillar via double lattice collapse and
  expansion},\ }\href {https://doi.org/10.1021/acs.nanolett.1c01750} {\bibfield
   {journal} {\bibinfo  {journal} {Nano Letters}\ }\textbf {\bibinfo {volume}
  {21}},\ \bibinfo {pages} {7913} (\bibinfo {year} {2021})},\ \bibinfo {note}
  {pMID: 34559544}\BibitemShut {NoStop}%
\bibitem [{\citenamefont {Gati}\ \emph {et~al.}(2016)\citenamefont {Gati},
  \citenamefont {Garst}, \citenamefont {Manna}, \citenamefont {Tutsch},
  \citenamefont {Wolf}, \citenamefont {Bartosch}, \citenamefont {Schubert},
  \citenamefont {Sasaki}, \citenamefont {Schlueter},\ and\ \citenamefont
  {Lang}}]{Gati2016}%
  \BibitemOpen
  \bibfield  {author} {\bibinfo {author} {\bibfnamefont {E.}~\bibnamefont
  {Gati}}, \bibinfo {author} {\bibfnamefont {M.}~\bibnamefont {Garst}},
  \bibinfo {author} {\bibfnamefont {R.~S.}\ \bibnamefont {Manna}}, \bibinfo
  {author} {\bibfnamefont {U.}~\bibnamefont {Tutsch}}, \bibinfo {author}
  {\bibfnamefont {B.}~\bibnamefont {Wolf}}, \bibinfo {author} {\bibfnamefont
  {L.}~\bibnamefont {Bartosch}}, \bibinfo {author} {\bibfnamefont
  {H.}~\bibnamefont {Schubert}}, \bibinfo {author} {\bibfnamefont
  {T.}~\bibnamefont {Sasaki}}, \bibinfo {author} {\bibfnamefont {J.~A.}\
  \bibnamefont {Schlueter}},\ and\ \bibinfo {author} {\bibfnamefont
  {M.}~\bibnamefont {Lang}},\ }\bibfield  {title} {\bibinfo {title} {Breakdown
  of hooke’s law of elasticity at the mott critical endpoint in an organic
  conductor},\ }\href {https://doi.org/10.1126/sciadv.1601646} {\bibfield
  {journal} {\bibinfo  {journal} {Science Advances}\ }\textbf {\bibinfo
  {volume} {2}},\ \bibinfo {pages} {e1601646} (\bibinfo {year}
  {2016})}\BibitemShut {NoStop}%
\bibitem [{\citenamefont {Peters}\ \emph {et~al.}(2023)\citenamefont {Peters},
  \citenamefont {Kliemt}, \citenamefont {Ocker}, \citenamefont {Wolf},
  \citenamefont {Puphal}, \citenamefont {Le~Tacon}, \citenamefont {Merz},
  \citenamefont {Lang},\ and\ \citenamefont {Krellner}}]{peters2023valence}%
  \BibitemOpen
  \bibfield  {author} {\bibinfo {author} {\bibfnamefont {M.}~\bibnamefont
  {Peters}}, \bibinfo {author} {\bibfnamefont {K.}~\bibnamefont {Kliemt}},
  \bibinfo {author} {\bibfnamefont {M.}~\bibnamefont {Ocker}}, \bibinfo
  {author} {\bibfnamefont {B.}~\bibnamefont {Wolf}}, \bibinfo {author}
  {\bibfnamefont {P.}~\bibnamefont {Puphal}}, \bibinfo {author} {\bibfnamefont
  {M.}~\bibnamefont {Le~Tacon}}, \bibinfo {author} {\bibfnamefont
  {M.}~\bibnamefont {Merz}}, \bibinfo {author} {\bibfnamefont {M.}~\bibnamefont
  {Lang}},\ and\ \bibinfo {author} {\bibfnamefont {C.}~\bibnamefont
  {Krellner}},\ }\bibfield  {title} {\bibinfo {title} {{F}rom valence
  fluctuations to long-range magnetic order in
  {E}u{P}d$_2$({S}i$_{1-x}${G}e$_x$)$_2$ single crystals},\ }\href
  {https://doi.org/10.1103/PhysRevMaterials.7.064405} {\bibfield  {journal}
  {\bibinfo  {journal} {Physical Review Materials}\ }\textbf {\bibinfo {volume}
  {7}},\ \bibinfo {pages} {064405} (\bibinfo {year} {2023})}\BibitemShut
  {NoStop}%
\bibitem [{\citenamefont {Wolf}\ \emph
  {et~al.}(2023{\natexlab{a}})\citenamefont {Wolf}, \citenamefont {Spathelf},
  \citenamefont {Zimmermann}, \citenamefont {Lundbeck}, \citenamefont {Peters},
  \citenamefont {Kliemt}, \citenamefont {Krellner},\ and\ \citenamefont
  {Lang}}]{Wolf2022}%
  \BibitemOpen
  \bibfield  {author} {\bibinfo {author} {\bibfnamefont {B.}~\bibnamefont
  {Wolf}}, \bibinfo {author} {\bibfnamefont {F.}~\bibnamefont {Spathelf}},
  \bibinfo {author} {\bibfnamefont {J.}~\bibnamefont {Zimmermann}}, \bibinfo
  {author} {\bibfnamefont {T.}~\bibnamefont {Lundbeck}}, \bibinfo {author}
  {\bibfnamefont {M.}~\bibnamefont {Peters}}, \bibinfo {author} {\bibfnamefont
  {K.}~\bibnamefont {Kliemt}}, \bibinfo {author} {\bibfnamefont
  {C.}~\bibnamefont {Krellner}},\ and\ \bibinfo {author} {\bibfnamefont
  {M.}~\bibnamefont {Lang}},\ }\bibfield  {title} {\bibinfo {title} {{From
  magnetic order to valence-change crossover in EuPd$_2$(Si$_{1-x}$Ge$_x$)$_2$
  using He-gas pressure}},\ }\href
  {https://doi.org/10.21468/SciPostPhysProc.11.022} {\bibfield  {journal}
  {\bibinfo  {journal} {SciPost Phys. Proc.}\ ,\ \bibinfo {pages} {022}}
  (\bibinfo {year} {2023}{\natexlab{a}})}\BibitemShut {NoStop}%
\bibitem [{\citenamefont {Wolf}\ \emph
  {et~al.}(2023{\natexlab{b}})\citenamefont {Wolf}, \citenamefont {Lundbeck},
  \citenamefont {Zimmermann}, \citenamefont {Peters}, \citenamefont {Kliemt},
  \citenamefont {Krellner},\ and\ \citenamefont {Lang}}]{Wolf2023}%
  \BibitemOpen
  \bibfield  {author} {\bibinfo {author} {\bibfnamefont {B.}~\bibnamefont
  {Wolf}}, \bibinfo {author} {\bibfnamefont {T.}~\bibnamefont {Lundbeck}},
  \bibinfo {author} {\bibfnamefont {J.}~\bibnamefont {Zimmermann}}, \bibinfo
  {author} {\bibfnamefont {M.}~\bibnamefont {Peters}}, \bibinfo {author}
  {\bibfnamefont {K.}~\bibnamefont {Kliemt}}, \bibinfo {author} {\bibfnamefont
  {C.}~\bibnamefont {Krellner}},\ and\ \bibinfo {author} {\bibfnamefont
  {M.}~\bibnamefont {Lang}},\ }\bibfield  {title} {\bibinfo {title} {{P}ressure
  study on the interplay between magnetic order and valence crossover in
  {${\mathrm{EuPd}}_{2}{({\mathrm{Si}}_{1\ensuremath{-}x}{\mathrm{Ge}}_{x})}_{2}$}},\
  }\href {https://doi.org/10.1103/PhysRevB.107.245147} {\bibfield  {journal}
  {\bibinfo  {journal} {Phys. Rev. B}\ }\textbf {\bibinfo {volume} {107}},\
  \bibinfo {pages} {245147} (\bibinfo {year} {2023}{\natexlab{b}})}\BibitemShut
  {NoStop}%
\bibitem [{\citenamefont {Mimura}\ \emph {et~al.}(2011)\citenamefont {Mimura},
  \citenamefont {Uozumi}, \citenamefont {Ishizu}, \citenamefont {Motonami},
  \citenamefont {Sato}, \citenamefont {Utsumi}, \citenamefont {Ueda},
  \citenamefont {Mitsuda}, \citenamefont {Shimada}, \citenamefont {Taguchi}
  \emph {et~al.}}]{mimura2011temperature}%
  \BibitemOpen
  \bibfield  {author} {\bibinfo {author} {\bibfnamefont {K.}~\bibnamefont
  {Mimura}}, \bibinfo {author} {\bibfnamefont {T.}~\bibnamefont {Uozumi}},
  \bibinfo {author} {\bibfnamefont {T.}~\bibnamefont {Ishizu}}, \bibinfo
  {author} {\bibfnamefont {S.}~\bibnamefont {Motonami}}, \bibinfo {author}
  {\bibfnamefont {H.}~\bibnamefont {Sato}}, \bibinfo {author} {\bibfnamefont
  {Y.}~\bibnamefont {Utsumi}}, \bibinfo {author} {\bibfnamefont
  {S.}~\bibnamefont {Ueda}}, \bibinfo {author} {\bibfnamefont {A.}~\bibnamefont
  {Mitsuda}}, \bibinfo {author} {\bibfnamefont {K.}~\bibnamefont {Shimada}},
  \bibinfo {author} {\bibfnamefont {Y.}~\bibnamefont {Taguchi}}, \emph
  {et~al.},\ }\bibfield  {title} {\bibinfo {title} {Temperature-induced valence
  transition of eupd2si2 studied by hard x-ray photoelectron spectroscopy},\
  }\href {https://doi.org/10.1143/JJAP.50.05FD03} {\bibfield  {journal}
  {\bibinfo  {journal} {Japanese journal of applied physics}\ }\textbf
  {\bibinfo {volume} {50}},\ \bibinfo {pages} {05FD03} (\bibinfo {year}
  {2011})}\BibitemShut {NoStop}%
\bibitem [{\citenamefont {Song}\ \emph {et~al.}(2023)\citenamefont {Song},
  \citenamefont {Schulz}, \citenamefont {Kliemt}, \citenamefont {Krellner},\
  and\ \citenamefont {Valent{\'\i}}}]{song2023microscopic}%
  \BibitemOpen
  \bibfield  {author} {\bibinfo {author} {\bibfnamefont {Y.-J.}\ \bibnamefont
  {Song}}, \bibinfo {author} {\bibfnamefont {S.}~\bibnamefont {Schulz}},
  \bibinfo {author} {\bibfnamefont {K.}~\bibnamefont {Kliemt}}, \bibinfo
  {author} {\bibfnamefont {C.}~\bibnamefont {Krellner}},\ and\ \bibinfo
  {author} {\bibfnamefont {R.}~\bibnamefont {Valent{\'\i}}},\ }\bibfield
  {title} {\bibinfo {title} {{M}icroscopic analysis of the valence transition
  in tetragonal {E}u{P}d$_2${S}i$_2$},\ }\href
  {https://doi.org/10.1103/PhysRevB.107.075149} {\bibfield  {journal} {\bibinfo
   {journal} {Physical Review B}\ }\textbf {\bibinfo {volume} {107}},\ \bibinfo
  {pages} {075149} (\bibinfo {year} {2023})}\BibitemShut {NoStop}%
\bibitem [{\citenamefont {Kliemt}\ \emph {et~al.}(2022)\citenamefont {Kliemt},
  \citenamefont {Peters}, \citenamefont {Reiser}, \citenamefont {Ocker},
  \citenamefont {Walther}, \citenamefont {Tran}, \citenamefont {Cho},
  \citenamefont {Merz}, \citenamefont {Haghighirad}, \citenamefont {Hezel}
  \emph {et~al.}}]{kliemt2022influence}%
  \BibitemOpen
  \bibfield  {author} {\bibinfo {author} {\bibfnamefont {K.}~\bibnamefont
  {Kliemt}}, \bibinfo {author} {\bibfnamefont {M.}~\bibnamefont {Peters}},
  \bibinfo {author} {\bibfnamefont {I.}~\bibnamefont {Reiser}}, \bibinfo
  {author} {\bibfnamefont {M.}~\bibnamefont {Ocker}}, \bibinfo {author}
  {\bibfnamefont {F.}~\bibnamefont {Walther}}, \bibinfo {author} {\bibfnamefont
  {D.-M.}\ \bibnamefont {Tran}}, \bibinfo {author} {\bibfnamefont
  {E.}~\bibnamefont {Cho}}, \bibinfo {author} {\bibfnamefont {M.}~\bibnamefont
  {Merz}}, \bibinfo {author} {\bibfnamefont {A.~A.}\ \bibnamefont
  {Haghighirad}}, \bibinfo {author} {\bibfnamefont {D.~C.}\ \bibnamefont
  {Hezel}}, \emph {et~al.},\ }\bibfield  {title} {\bibinfo {title} {{I}nfluence
  of the {P}d-{S}i {R}atio on the {V}alence {T}ransition in
  {E}u{P}d$_2${S}i$_2$ {S}ingle {C}rystals},\ }\href
  {https://doi.org/10.1021/acs.cgd.2c00475} {\bibfield  {journal} {\bibinfo
  {journal} {Crystal Growth \& Design}\ }\textbf {\bibinfo {volume} {22}},\
  \bibinfo {pages} {5399} (\bibinfo {year} {2022})}\BibitemShut {NoStop}%
\bibitem [{\citenamefont {Felner}\ and\ \citenamefont
  {Nowik}(1986)}]{felner1986first}%
  \BibitemOpen
  \bibfield  {author} {\bibinfo {author} {\bibfnamefont {I.}~\bibnamefont
  {Felner}}\ and\ \bibinfo {author} {\bibfnamefont {I.}~\bibnamefont {Nowik}},\
  }\bibfield  {title} {\bibinfo {title} {{First-order valence phase transition
  in cubic Yb$_x$In$_{1-x}$Cu$_2$}},\ }\href
  {https://doi.org/https://doi.org/10.1103/PhysRevB.33.617} {\bibfield
  {journal} {\bibinfo  {journal} {Physical Review B}\ }\textbf {\bibinfo
  {volume} {33}},\ \bibinfo {pages} {617} (\bibinfo {year} {1986})}\BibitemShut
  {NoStop}%
\bibitem [{\citenamefont {Sarrao}\ \emph
  {et~al.}(1996{\natexlab{a}})\citenamefont {Sarrao}, \citenamefont {Immer},
  \citenamefont {Benton}, \citenamefont {Fisk}, \citenamefont {Lawrence},
  \citenamefont {Mandrus},\ and\ \citenamefont
  {Thompson}}]{sarrao1996evolution}%
  \BibitemOpen
  \bibfield  {author} {\bibinfo {author} {\bibfnamefont {J.}~\bibnamefont
  {Sarrao}}, \bibinfo {author} {\bibfnamefont {C.}~\bibnamefont {Immer}},
  \bibinfo {author} {\bibfnamefont {C.}~\bibnamefont {Benton}}, \bibinfo
  {author} {\bibfnamefont {Z.}~\bibnamefont {Fisk}}, \bibinfo {author}
  {\bibfnamefont {J.}~\bibnamefont {Lawrence}}, \bibinfo {author}
  {\bibfnamefont {D.}~\bibnamefont {Mandrus}},\ and\ \bibinfo {author}
  {\bibfnamefont {J.}~\bibnamefont {Thompson}},\ }\bibfield  {title} {\bibinfo
  {title} {{Evolution from first-order valence transition to heavy-fermion
  behavior in YbIn$_{1- x}$Ag$_x$Cu$_4$}},\ }\href
  {https://doi.org/https://doi.org/10.1103/PhysRevB.54.12207} {\bibfield
  {journal} {\bibinfo  {journal} {Physical Review B}\ }\textbf {\bibinfo
  {volume} {54}},\ \bibinfo {pages} {12207} (\bibinfo {year}
  {1996}{\natexlab{a}})}\BibitemShut {NoStop}%
\bibitem [{\citenamefont {Sarrao}\ \emph {et~al.}(1998)\citenamefont {Sarrao},
  \citenamefont {Ramirez}, \citenamefont {Darling}, \citenamefont {Freibert},
  \citenamefont {Migliori}, \citenamefont {Immer}, \citenamefont {Fisk},\ and\
  \citenamefont {Uwatoko}}]{sarrao1998thermodynamics}%
  \BibitemOpen
  \bibfield  {author} {\bibinfo {author} {\bibfnamefont {J.}~\bibnamefont
  {Sarrao}}, \bibinfo {author} {\bibfnamefont {A.}~\bibnamefont {Ramirez}},
  \bibinfo {author} {\bibfnamefont {T.}~\bibnamefont {Darling}}, \bibinfo
  {author} {\bibfnamefont {F.}~\bibnamefont {Freibert}}, \bibinfo {author}
  {\bibfnamefont {A.}~\bibnamefont {Migliori}}, \bibinfo {author}
  {\bibfnamefont {C.}~\bibnamefont {Immer}}, \bibinfo {author} {\bibfnamefont
  {Z.}~\bibnamefont {Fisk}},\ and\ \bibinfo {author} {\bibfnamefont
  {Y.}~\bibnamefont {Uwatoko}},\ }\bibfield  {title} {\bibinfo {title}
  {{Thermodynamics of the first-order valence transition in YbInCu$_4$}},\
  }\href {https://doi.org/https://doi.org/10.1103/PhysRevB.58.409} {\bibfield
  {journal} {\bibinfo  {journal} {Physical Review B}\ }\textbf {\bibinfo
  {volume} {58}},\ \bibinfo {pages} {409} (\bibinfo {year} {1998})}\BibitemShut
  {NoStop}%
\bibitem [{\citenamefont {Katori}\ \emph {et~al.}(1994)\citenamefont {Katori},
  \citenamefont {Goto},\ and\ \citenamefont {Yoshimura}}]{katori1994field}%
  \BibitemOpen
  \bibfield  {author} {\bibinfo {author} {\bibfnamefont {H.~A.}\ \bibnamefont
  {Katori}}, \bibinfo {author} {\bibfnamefont {T.}~\bibnamefont {Goto}},\ and\
  \bibinfo {author} {\bibfnamefont {K.}~\bibnamefont {Yoshimura}},\ }\bibfield
  {title} {\bibinfo {title} {{Field-induced metamagnetic transition in valence
  fluctuating compound YbIn$_{1- x}$Ag$_x$Cu$_4$}},\ }\href
  {https://doi.org/https://doi.org/10.1016/0921-4526(94)91073-1} {\bibfield
  {journal} {\bibinfo  {journal} {Physica B: Condensed Matter}\ }\textbf
  {\bibinfo {volume} {201}},\ \bibinfo {pages} {159} (\bibinfo {year}
  {1994})}\BibitemShut {NoStop}%
\bibitem [{\citenamefont {L{\"o}ffert}\ \emph {et~al.}(1999)\citenamefont
  {L{\"o}ffert}, \citenamefont {Hautsch}, \citenamefont {Ritter},\ and\
  \citenamefont {Assmus}}]{loffert1999phase}%
  \BibitemOpen
  \bibfield  {author} {\bibinfo {author} {\bibfnamefont {A.}~\bibnamefont
  {L{\"o}ffert}}, \bibinfo {author} {\bibfnamefont {S.}~\bibnamefont
  {Hautsch}}, \bibinfo {author} {\bibfnamefont {F.}~\bibnamefont {Ritter}},\
  and\ \bibinfo {author} {\bibfnamefont {W.}~\bibnamefont {Assmus}},\
  }\bibfield  {title} {\bibinfo {title} {{The phase diagram of YbInCu$_4$}},\
  }\href {https://doi.org/https://doi.org/10.1016/S0921-4526(98)01067-9}
  {\bibfield  {journal} {\bibinfo  {journal} {Physica B: Condensed Matter}\
  }\textbf {\bibinfo {volume} {259}},\ \bibinfo {pages} {134} (\bibinfo {year}
  {1999})}\BibitemShut {NoStop}%
\bibitem [{\citenamefont {Hiraoka}\ \emph {et~al.}(2007)\citenamefont
  {Hiraoka}, \citenamefont {Yabuta}, \citenamefont {Kojima}, \citenamefont
  {Oikawa},\ and\ \citenamefont {Kamiyama}}]{hiraoka2007neutron}%
  \BibitemOpen
  \bibfield  {author} {\bibinfo {author} {\bibfnamefont {K.}~\bibnamefont
  {Hiraoka}}, \bibinfo {author} {\bibfnamefont {H.}~\bibnamefont {Yabuta}},
  \bibinfo {author} {\bibfnamefont {K.}~\bibnamefont {Kojima}}, \bibinfo
  {author} {\bibfnamefont {K.}~\bibnamefont {Oikawa}},\ and\ \bibinfo {author}
  {\bibfnamefont {T.}~\bibnamefont {Kamiyama}},\ }\bibfield  {title} {\bibinfo
  {title} {{Neutron powder diffraction study of the site disorder in
  YbInCu$_4$}},\ }\href
  {https://doi.org/https://doi.org/10.1016/j.jmmm.2006.10.085} {\bibfield
  {journal} {\bibinfo  {journal} {Journal of magnetism and magnetic materials}\
  }\textbf {\bibinfo {volume} {310}},\ \bibinfo {pages} {380} (\bibinfo {year}
  {2007})}\BibitemShut {NoStop}%
\bibitem [{\citenamefont {Kojima}\ \emph {et~al.}(1992)\citenamefont {Kojima},
  \citenamefont {Yabuta},\ and\ \citenamefont {Hihara}}]{Kojima1992}%
  \BibitemOpen
  \bibfield  {author} {\bibinfo {author} {\bibfnamefont {K.}~\bibnamefont
  {Kojima}}, \bibinfo {author} {\bibfnamefont {H.}~\bibnamefont {Yabuta}},\
  and\ \bibinfo {author} {\bibfnamefont {T.}~\bibnamefont {Hihara}},\
  }\bibfield  {title} {\bibinfo {title} {{Magnetic and NMR study of valence
  phase transition in YbIn$_{1-x}T_x$Cu$_4$ ($T$ = Ag and Au)}},\ }\href
  {https://doi.org/https://doi.org/10.1016/0304-8853(92)90968-T} {\bibfield
  {journal} {\bibinfo  {journal} {Journal of Magnetism and Magnetic Materials}\
  }\textbf {\bibinfo {volume} {104-107}},\ \bibinfo {pages} {653} (\bibinfo
  {year} {1992})}\BibitemShut {NoStop}%
\bibitem [{\citenamefont {Sarrao}\ \emph
  {et~al.}(1996{\natexlab{b}})\citenamefont {Sarrao}, \citenamefont {Benton},
  \citenamefont {Fisk}, \citenamefont {Lawrence}, \citenamefont {Mandrus},\
  and\ \citenamefont {Thompson}}]{sarrao1996ybin1}%
  \BibitemOpen
  \bibfield  {author} {\bibinfo {author} {\bibfnamefont {J.}~\bibnamefont
  {Sarrao}}, \bibinfo {author} {\bibfnamefont {C.}~\bibnamefont {Benton}},
  \bibinfo {author} {\bibfnamefont {Z.}~\bibnamefont {Fisk}}, \bibinfo {author}
  {\bibfnamefont {J.}~\bibnamefont {Lawrence}}, \bibinfo {author}
  {\bibfnamefont {D.}~\bibnamefont {Mandrus}},\ and\ \bibinfo {author}
  {\bibfnamefont {J.}~\bibnamefont {Thompson}},\ }\bibfield  {title} {\bibinfo
  {title} {{YbIn$_{1- x}$Ag$_x$Cu$_4$: Crossover from first-order valence
  transition to heavy Fermion behavior}},\ }\href
  {https://doi.org/https://doi.org/10.1016/0921-4526(96)00124-X} {\bibfield
  {journal} {\bibinfo  {journal} {Physica B: Condensed Matter}\ }\textbf
  {\bibinfo {volume} {223}},\ \bibinfo {pages} {366} (\bibinfo {year}
  {1996}{\natexlab{b}})}\BibitemShut {NoStop}%
\bibitem [{\citenamefont {Severing}\ \emph {et~al.}(1990)\citenamefont
  {Severing}, \citenamefont {Murani}, \citenamefont {Thompson}, \citenamefont
  {Fisk},\ and\ \citenamefont {Loong}}]{severing1990neutron}%
  \BibitemOpen
  \bibfield  {author} {\bibinfo {author} {\bibfnamefont {A.}~\bibnamefont
  {Severing}}, \bibinfo {author} {\bibfnamefont {A.}~\bibnamefont {Murani}},
  \bibinfo {author} {\bibfnamefont {J.~D.}\ \bibnamefont {Thompson}}, \bibinfo
  {author} {\bibfnamefont {Z.}~\bibnamefont {Fisk}},\ and\ \bibinfo {author}
  {\bibfnamefont {C.-K.}\ \bibnamefont {Loong}},\ }\bibfield  {title} {\bibinfo
  {title} {Neutron scattering experiments on {Y}b{X}{C}u$_4$ and
  {E}r{X}{C}u$_4$ ({X}= {A}u, {P}d, and {A}g)},\ }\href
  {https://doi.org/10.1103/PhysRevB.41.1739} {\bibfield  {journal} {\bibinfo
  {journal} {Physical Review B}\ }\textbf {\bibinfo {volume} {41}},\ \bibinfo
  {pages} {1739} (\bibinfo {year} {1990})}\BibitemShut {NoStop}%
\bibitem [{\citenamefont {Casanova}\ \emph {et~al.}(1990)\citenamefont
  {Casanova}, \citenamefont {Jaccard}, \citenamefont {Marcenat}, \citenamefont
  {Hamdaoui},\ and\ \citenamefont {Besnus}}]{casanova1990thermoelectric}%
  \BibitemOpen
  \bibfield  {author} {\bibinfo {author} {\bibfnamefont {R.}~\bibnamefont
  {Casanova}}, \bibinfo {author} {\bibfnamefont {D.}~\bibnamefont {Jaccard}},
  \bibinfo {author} {\bibfnamefont {C.}~\bibnamefont {Marcenat}}, \bibinfo
  {author} {\bibfnamefont {N.}~\bibnamefont {Hamdaoui}},\ and\ \bibinfo
  {author} {\bibfnamefont {M.}~\bibnamefont {Besnus}},\ }\bibfield  {title}
  {\bibinfo {title} {{Thermoelectric power of YbMCu$_4$ (M= Ag, Au and Pd) and
  YbPd$_2$Si$_2$}},\ }\href {https://doi.org/10.1016/S0304-8853(10)80217-3}
  {\bibfield  {journal} {\bibinfo  {journal} {Journal of magnetism and magnetic
  materials}\ }\textbf {\bibinfo {volume} {90}},\ \bibinfo {pages} {587}
  (\bibinfo {year} {1990})}\BibitemShut {NoStop}%
\bibitem [{\citenamefont {Tkeuchi}\ \emph {et~al.}(2015)\citenamefont
  {Tkeuchi}, \citenamefont {Hirose}, \citenamefont {Tsunoda}, \citenamefont
  {Honda},\ and\ \citenamefont {Settai}}]{tkeuchi2015themal}%
  \BibitemOpen
  \bibfield  {author} {\bibinfo {author} {\bibfnamefont {T.}~\bibnamefont
  {Tkeuchi}}, \bibinfo {author} {\bibfnamefont {Y.}~\bibnamefont {Hirose}},
  \bibinfo {author} {\bibfnamefont {R.}~\bibnamefont {Tsunoda}}, \bibinfo
  {author} {\bibfnamefont {F.}~\bibnamefont {Honda}},\ and\ \bibinfo {author}
  {\bibfnamefont {R.}~\bibnamefont {Settai}},\ }\bibfield  {title} {\bibinfo
  {title} {{Thermal Expansion and Magnetostriction of YbAuCu$_4$}},\ }\href
  {https://doi.org/https://doi.org/10.1016/j.phpro.2015.12.057} {\bibfield
  {journal} {\bibinfo  {journal} {Physics Procedia}\ }\textbf {\bibinfo
  {volume} {75}},\ \bibinfo {pages} {460} (\bibinfo {year} {2015})}\BibitemShut
  {NoStop}%
\bibitem [{\citenamefont {Wada}\ and\ \citenamefont
  {Yamamoto}(2008)}]{wada2008non}%
  \BibitemOpen
  \bibfield  {author} {\bibinfo {author} {\bibfnamefont {S.}~\bibnamefont
  {Wada}}\ and\ \bibinfo {author} {\bibfnamefont {A.}~\bibnamefont
  {Yamamoto}},\ }\bibfield  {title} {\bibinfo {title} {{Non-Fermi liquid
  phenomena and intermediate valence in Yb-based compounds located close to the
  quantum critical point}},\ }\href
  {https://doi.org/https://doi.org/10.1016/j.physb.2007.10.250} {\bibfield
  {journal} {\bibinfo  {journal} {Physica B: Condensed Matter}\ }\textbf
  {\bibinfo {volume} {403}},\ \bibinfo {pages} {1202} (\bibinfo {year}
  {2008})}\BibitemShut {NoStop}%
\bibitem [{\citenamefont {Cornelius}\ \emph {et~al.}(1997)\citenamefont
  {Cornelius}, \citenamefont {Lawrence}, \citenamefont {Sarrao}, \citenamefont
  {Fisk}, \citenamefont {Hundley}, \citenamefont {Kwei}, \citenamefont
  {Thompson}, \citenamefont {Booth},\ and\ \citenamefont
  {Bridges}}]{Cornelius1997}%
  \BibitemOpen
  \bibfield  {author} {\bibinfo {author} {\bibfnamefont {A.~L.}\ \bibnamefont
  {Cornelius}}, \bibinfo {author} {\bibfnamefont {J.~M.}\ \bibnamefont
  {Lawrence}}, \bibinfo {author} {\bibfnamefont {J.~L.}\ \bibnamefont
  {Sarrao}}, \bibinfo {author} {\bibfnamefont {Z.}~\bibnamefont {Fisk}},
  \bibinfo {author} {\bibfnamefont {M.~F.}\ \bibnamefont {Hundley}}, \bibinfo
  {author} {\bibfnamefont {G.~H.}\ \bibnamefont {Kwei}}, \bibinfo {author}
  {\bibfnamefont {J.~D.}\ \bibnamefont {Thompson}}, \bibinfo {author}
  {\bibfnamefont {C.~H.}\ \bibnamefont {Booth}},\ and\ \bibinfo {author}
  {\bibfnamefont {F.}~\bibnamefont {Bridges}},\ }\bibfield  {title} {\bibinfo
  {title} {{Experimental studies of the phase transition in
  YbIn$_{1-x}$Ag$_x$Cu$_4$}},\ }\href
  {https://doi.org/10.1103/PhysRevB.56.7993} {\bibfield  {journal} {\bibinfo
  {journal} {Phys. Rev. B}\ }\textbf {\bibinfo {volume} {56}},\ \bibinfo
  {pages} {7993} (\bibinfo {year} {1997})}\BibitemShut {NoStop}%
\bibitem [{\citenamefont {Zacharias}\ \emph {et~al.}(2015)\citenamefont
  {Zacharias}, \citenamefont {Rosch},\ and\ \citenamefont
  {Garst}}]{Zacharias2015}%
  \BibitemOpen
  \bibfield  {author} {\bibinfo {author} {\bibfnamefont {M.}~\bibnamefont
  {Zacharias}}, \bibinfo {author} {\bibfnamefont {A.}~\bibnamefont {Rosch}},\
  and\ \bibinfo {author} {\bibfnamefont {M.}~\bibnamefont {Garst}},\ }\bibfield
   {title} {\bibinfo {title} {{C}ritical elasticity at zero and finite
  temperature},\ }\href {https://doi.org/10.1140/epjst/e2015-02444-5}
  {\bibfield  {journal} {\bibinfo  {journal} {The European Physical Journal
  Special Topics}\ }\textbf {\bibinfo {volume} {224}},\ \bibinfo {pages} {1021}
  (\bibinfo {year} {2015})}\BibitemShut {NoStop}%
\bibitem [{\citenamefont {Paul C.~Canfield}\ and\ \citenamefont
  {Jo}(2016)}]{Canfield2016}%
  \BibitemOpen
  \bibfield  {author} {\bibinfo {author} {\bibfnamefont {U.~S.~K.}\
  \bibnamefont {Paul C.~Canfield}, \bibfnamefont {Tai~Kong}}\ and\ \bibinfo
  {author} {\bibfnamefont {N.~H.}\ \bibnamefont {Jo}},\ }\bibfield  {title}
  {\bibinfo {title} {Use of frit-disc crucibles for routine and exploratory
  solution growth of single crystalline samples},\ }\href
  {https://doi.org/10.1080/14786435.2015.1122248} {\bibfield  {journal}
  {\bibinfo  {journal} {Philosophical Magazine}\ }\textbf {\bibinfo {volume}
  {96}},\ \bibinfo {pages} {84} (\bibinfo {year} {2016})}\BibitemShut {NoStop}%
\bibitem [{\citenamefont {Toby}\ and\ \citenamefont
  {Von~Dreele}(2013)}]{toby2013gsas}%
  \BibitemOpen
  \bibfield  {author} {\bibinfo {author} {\bibfnamefont {B.~H.}\ \bibnamefont
  {Toby}}\ and\ \bibinfo {author} {\bibfnamefont {R.~B.}\ \bibnamefont
  {Von~Dreele}},\ }\bibfield  {title} {\bibinfo {title} {{GSAS-II}: the genesis
  of a modern open-source all purpose crystallography software package},\
  }\href {https://doi.org/10.1107/S0021889813003531} {\bibfield  {journal}
  {\bibinfo  {journal} {Journal of Applied Crystallography}\ }\textbf {\bibinfo
  {volume} {46}},\ \bibinfo {pages} {544} (\bibinfo {year} {2013})}\BibitemShut
  {NoStop}%
\bibitem [{\citenamefont {Ocker}\ \emph {et~al.}(2025)\citenamefont {Ocker},
  \citenamefont {Ghebretinsae}, \citenamefont {Zimmermann}, \citenamefont
  {Würtele}, \citenamefont {Wolf}, \citenamefont {Virovets}, \citenamefont
  {Lang}, \citenamefont {Kliemt},\ and\ \citenamefont
  {Krellner}}]{supplementalinfo_YbInCu4_2024}%
  \BibitemOpen
  \bibfield  {author} {\bibinfo {author} {\bibfnamefont {M.}~\bibnamefont
  {Ocker}}, \bibinfo {author} {\bibfnamefont {B.}~\bibnamefont {Ghebretinsae}},
  \bibinfo {author} {\bibfnamefont {J.-N.}\ \bibnamefont {Zimmermann}},
  \bibinfo {author} {\bibfnamefont {S.}~\bibnamefont {Würtele}}, \bibinfo
  {author} {\bibfnamefont {B.}~\bibnamefont {Wolf}}, \bibinfo {author}
  {\bibfnamefont {A.}~\bibnamefont {Virovets}}, \bibinfo {author}
  {\bibfnamefont {M.}~\bibnamefont {Lang}}, \bibinfo {author} {\bibfnamefont
  {K.}~\bibnamefont {Kliemt}},\ and\ \bibinfo {author} {\bibfnamefont
  {C.}~\bibnamefont {Krellner}},\ }\bibfield  {title} {\bibinfo {title} {{See
  supplemental material at XXXXXXXX for additional data of the heat capacity,
  results of the SC-XRD, Laue analysis, and magnetic susceptibility.}},\
  }\href@noop {} {\  (\bibinfo {year} {2025})}\BibitemShut {NoStop}%
\bibitem [{\citenamefont {Lashley}\ \emph {et~al.}(2003)\citenamefont
  {Lashley}, \citenamefont {Hundley}, \citenamefont {Migliori}, \citenamefont
  {Sarrao}, \citenamefont {Pagliuso}, \citenamefont {Darling}, \citenamefont
  {Jaime}, \citenamefont {Cooley}, \citenamefont {Hults}, \citenamefont
  {Morales} \emph {et~al.}}]{lashley2003critical}%
  \BibitemOpen
  \bibfield  {author} {\bibinfo {author} {\bibfnamefont {J.}~\bibnamefont
  {Lashley}}, \bibinfo {author} {\bibfnamefont {M.}~\bibnamefont {Hundley}},
  \bibinfo {author} {\bibfnamefont {A.}~\bibnamefont {Migliori}}, \bibinfo
  {author} {\bibfnamefont {J.}~\bibnamefont {Sarrao}}, \bibinfo {author}
  {\bibfnamefont {P.}~\bibnamefont {Pagliuso}}, \bibinfo {author}
  {\bibfnamefont {T.}~\bibnamefont {Darling}}, \bibinfo {author} {\bibfnamefont
  {M.}~\bibnamefont {Jaime}}, \bibinfo {author} {\bibfnamefont
  {J.}~\bibnamefont {Cooley}}, \bibinfo {author} {\bibfnamefont
  {W.}~\bibnamefont {Hults}}, \bibinfo {author} {\bibfnamefont
  {L.}~\bibnamefont {Morales}}, \emph {et~al.},\ }\bibfield  {title} {\bibinfo
  {title} {{C}ritical examination of heat capacity measurements made on a
  {Q}uantum {D}esign physical property measurement system},\ }\href
  {https://doi.org/10.1016/S0011-2275(03)00092-4} {\bibfield  {journal}
  {\bibinfo  {journal} {Cryogenics}\ }\textbf {\bibinfo {volume} {43}},\
  \bibinfo {pages} {369} (\bibinfo {year} {2003})}\BibitemShut {NoStop}%
\bibitem [{\citenamefont {L{\"u}thi}\ \emph {et~al.}(1994)\citenamefont
  {L{\"u}thi}, \citenamefont {Bruls}, \citenamefont {Thalmeier}, \citenamefont
  {Wolf}, \citenamefont {Finsterbusch},\ and\ \citenamefont
  {Kouroudis}}]{luthi1994electron}%
  \BibitemOpen
  \bibfield  {author} {\bibinfo {author} {\bibfnamefont {B.}~\bibnamefont
  {L{\"u}thi}}, \bibinfo {author} {\bibfnamefont {G.}~\bibnamefont {Bruls}},
  \bibinfo {author} {\bibfnamefont {P.}~\bibnamefont {Thalmeier}}, \bibinfo
  {author} {\bibfnamefont {B.}~\bibnamefont {Wolf}}, \bibinfo {author}
  {\bibfnamefont {D.}~\bibnamefont {Finsterbusch}},\ and\ \bibinfo {author}
  {\bibfnamefont {I.}~\bibnamefont {Kouroudis}},\ }\bibfield  {title} {\bibinfo
  {title} {Electron-phonon effects in heavy fermion systems},\ }\href
  {https://doi.org/10.1007/BF00754941} {\bibfield  {journal} {\bibinfo
  {journal} {Journal of Low Temperature Physics}\ }\textbf {\bibinfo {volume}
  {95}},\ \bibinfo {pages} {257} (\bibinfo {year} {1994})}\BibitemShut
  {NoStop}%
\bibitem [{\citenamefont {Liu}\ \emph {et~al.}(2002)\citenamefont {Liu},
  \citenamefont {Cui}, \citenamefont {Ishida}, \citenamefont {Liu},
  \citenamefont {Wang}, \citenamefont {Ohnuma}, \citenamefont {Kainuma},\ and\
  \citenamefont {Jin}}]{liu2002thermodynamic}%
  \BibitemOpen
  \bibfield  {author} {\bibinfo {author} {\bibfnamefont {H.}~\bibnamefont
  {Liu}}, \bibinfo {author} {\bibfnamefont {Y.}~\bibnamefont {Cui}}, \bibinfo
  {author} {\bibfnamefont {K.}~\bibnamefont {Ishida}}, \bibinfo {author}
  {\bibfnamefont {X.}~\bibnamefont {Liu}}, \bibinfo {author} {\bibfnamefont
  {C.}~\bibnamefont {Wang}}, \bibinfo {author} {\bibfnamefont {I.}~\bibnamefont
  {Ohnuma}}, \bibinfo {author} {\bibfnamefont {R.}~\bibnamefont {Kainuma}},\
  and\ \bibinfo {author} {\bibfnamefont {Z.}~\bibnamefont {Jin}},\ }\bibfield
  {title} {\bibinfo {title} {{Thermodynamic assessment of the Cu-In binary
  system}},\ }\href
  {https://doi.org/https://doi.org/10.1361/105497102770331352} {\bibfield
  {journal} {\bibinfo  {journal} {Journal of Phase Equilibria}\ }\textbf
  {\bibinfo {volume} {23}},\ \bibinfo {pages} {409} (\bibinfo {year}
  {2002})}\BibitemShut {NoStop}%
\bibitem [{\citenamefont {Fischbach}\ \emph {et~al.}(1998)\citenamefont
  {Fischbach}, \citenamefont {L{\"o}ffert}, \citenamefont {Ritter},\ and\
  \citenamefont {Assmus}}]{fischbach1998thermoanalytical}%
  \BibitemOpen
  \bibfield  {author} {\bibinfo {author} {\bibfnamefont {E.}~\bibnamefont
  {Fischbach}}, \bibinfo {author} {\bibfnamefont {A.}~\bibnamefont
  {L{\"o}ffert}}, \bibinfo {author} {\bibfnamefont {F.}~\bibnamefont
  {Ritter}},\ and\ \bibinfo {author} {\bibfnamefont {W.}~\bibnamefont
  {Assmus}},\ }\bibfield  {title} {\bibinfo {title} {{Thermoanalytical
  Investigations to Understand the Dependence Between the Growth Method and
  Crystal Properties of Valence Changing “YbInCu$_4$”}},\ }\href
  {https://doi.org/https://doi.org/10.1002/(SICI)1521-4079(1998)33:2%3C267::AID-CRAT267%3E3.0.CO;2-S}
  {\bibfield  {journal} {\bibinfo  {journal} {Crystal Research and Technology:
  Journal of Experimental and Industrial Crystallography}\ }\textbf {\bibinfo
  {volume} {33}},\ \bibinfo {pages} {267} (\bibinfo {year} {1998})}\BibitemShut
  {NoStop}%
\bibitem [{\citenamefont {Moriyoshi}\ \emph {et~al.}(2003)\citenamefont
  {Moriyoshi}, \citenamefont {Shimomura}, \citenamefont {Itoh}, \citenamefont
  {Kojima},\ and\ \citenamefont {Hiraoka}}]{moriyoshi2003crystal}%
  \BibitemOpen
  \bibfield  {author} {\bibinfo {author} {\bibfnamefont {C.}~\bibnamefont
  {Moriyoshi}}, \bibinfo {author} {\bibfnamefont {S.}~\bibnamefont
  {Shimomura}}, \bibinfo {author} {\bibfnamefont {K.}~\bibnamefont {Itoh}},
  \bibinfo {author} {\bibfnamefont {K.}~\bibnamefont {Kojima}},\ and\ \bibinfo
  {author} {\bibfnamefont {K.}~\bibnamefont {Hiraoka}},\ }\bibfield  {title}
  {\bibinfo {title} {{Crystal structure and valence transition temperature of
  YbInCu$_4$ single crystals}},\ }\href
  {https://doi.org/https://doi.org/10.1016/S0304-8853(02)01324-0} {\bibfield
  {journal} {\bibinfo  {journal} {Journal of magnetism and magnetic materials}\
  }\textbf {\bibinfo {volume} {260}},\ \bibinfo {pages} {206} (\bibinfo {year}
  {2003})}\BibitemShut {NoStop}%
\bibitem [{\citenamefont {Lawrence}\ \emph {et~al.}(1996)\citenamefont
  {Lawrence}, \citenamefont {Kwei}, \citenamefont {Sarrao}, \citenamefont
  {Fisk}, \citenamefont {Mandrus},\ and\ \citenamefont
  {Thompson}}]{lawrence1996structure}%
  \BibitemOpen
  \bibfield  {author} {\bibinfo {author} {\bibfnamefont {J.}~\bibnamefont
  {Lawrence}}, \bibinfo {author} {\bibfnamefont {G.}~\bibnamefont {Kwei}},
  \bibinfo {author} {\bibfnamefont {J.}~\bibnamefont {Sarrao}}, \bibinfo
  {author} {\bibfnamefont {Z.}~\bibnamefont {Fisk}}, \bibinfo {author}
  {\bibfnamefont {D.}~\bibnamefont {Mandrus}},\ and\ \bibinfo {author}
  {\bibfnamefont {J.}~\bibnamefont {Thompson}},\ }\bibfield  {title} {\bibinfo
  {title} {{Structure and disorder in YbInCu$_4$}},\ }\href
  {https://doi.org/https://doi.org/10.1103/PhysRevB.54.6011} {\bibfield
  {journal} {\bibinfo  {journal} {Physical Review B}\ }\textbf {\bibinfo
  {volume} {54}},\ \bibinfo {pages} {6011} (\bibinfo {year}
  {1996})}\BibitemShut {NoStop}%
\bibitem [{\citenamefont {Zherlitsyn}\ \emph {et~al.}(1999)\citenamefont
  {Zherlitsyn}, \citenamefont {L\"uthi}, \citenamefont {Wolf}, \citenamefont
  {Sarrao}, \citenamefont {Fisk},\ and\ \citenamefont
  {Zlati\ifmmode~\acute{c}\else \'{c}\fi{}}}]{Zherlitsyn1999}%
  \BibitemOpen
  \bibfield  {author} {\bibinfo {author} {\bibfnamefont {S.}~\bibnamefont
  {Zherlitsyn}}, \bibinfo {author} {\bibfnamefont {B.}~\bibnamefont {L\"uthi}},
  \bibinfo {author} {\bibfnamefont {B.}~\bibnamefont {Wolf}}, \bibinfo {author}
  {\bibfnamefont {J.~L.}\ \bibnamefont {Sarrao}}, \bibinfo {author}
  {\bibfnamefont {Z.}~\bibnamefont {Fisk}},\ and\ \bibinfo {author}
  {\bibfnamefont {V.}~\bibnamefont {Zlati\ifmmode~\acute{c}\else \'{c}\fi{}}},\
  }\bibfield  {title} {\bibinfo {title} {{U}ltrasonic study of the
  mixed-valence system {Y}b{I}n$_{1-x}${A}g$_x${C}u$_4$},\ }\href
  {https://doi.org/10.1103/PhysRevB.60.3148} {\bibfield  {journal} {\bibinfo
  {journal} {Phys. Rev. B}\ }\textbf {\bibinfo {volume} {60}},\ \bibinfo
  {pages} {3148} (\bibinfo {year} {1999})}\BibitemShut {NoStop}%
\end{thebibliography}%
\end{document}